\documentclass[acmtog,screen,nonacm]{acmart}

\usepackage{booktabs} 

\usepackage[ruled]{algorithm2e} 

\SetAlFnt{\small}
\SetAlCapFnt{\small}
\SetAlCapNameFnt{\small}
\SetAlCapHSkip{0pt}

\acmJournal{TOG}

\makeatletter
\let\@authorsaddresses\@empty
\makeatother

\usepackage[inline]{enumitem} 
\usepackage{color}
\usepackage{multirow}
\usepackage[makeroom]{cancel}
\usepackage{bm}

\definecolor{turquoise}{cmyk}{0.65,0,0.1,0.3}
\definecolor{purple}{rgb}{0.65,0,0.65}
\definecolor{dark_purple}{rgb}{0.5,0,0.5}
\definecolor{dark_green}{rgb}{0, 0.5, 0}
\definecolor{orange}{rgb}{0.8, 0.6, 0.2}
\definecolor{red}{rgb}{0.8, 0.2, 0.2}
\definecolor{darkred}{rgb}{0.6, 0.1, 0.05}
\definecolor{blueish}{rgb}{0.0, 0.3, .6}
\definecolor{light_gray}{rgb}{0.7, 0.7, .7}
\definecolor{pink}{rgb}{1, 0, 1}
\definecolor{greyblue}{rgb}{0.25, 0.25, 1}

\newcommand{\rz}[1]{{\color{blue}#1}}

\newcommand{\qm}[1]{{\color{red}#1}}

\begin{document}
\title{ART-DECO: \underline{Ar}bitrary \underline{T}ext Guidance for 3D \underline{De}tailizer \underline{Co}nstruction}

\author{Qimin Chen}
\affiliation{%
 \institution{Simon Fraser University}
 \country{Canada}}
\email{qca43@sfu.ca}

\author{Yuezhi Yang}
\affiliation{%
 \institution{University of Texas at Austin}
 \country{USA}
}
\email{yyuezhi123@gmail.com}

\author{Wang Yifan}
\affiliation{%
\institution{Adobe Research}
\country{USA}}
\email{yifwang@adobe.com}

\author{Vladimir Kim}
\affiliation{%
 \institution{Adobe Research}
 \country{USA}
}
\email{vova.g.kim@gmail.com}

\author{Siddhartha Chaudhuri}
\affiliation{%
 \institution{Adobe Research}
 \country{USA}}
\email{siddhartha.chaudhuri@gmail.com}

\author{Hao Zhang}
\affiliation{%
 \institution{Simon Fraser University}
 \country{Canada}}
\email{haoz@sfu.ca}

\author{Zhiqin Chen}
\affiliation{%
 \institution{Adobe Research}
 \country{USA}}
\email{chenzhiqin142857@gmail.com}

\begin{teaserfigure}
  \begin{picture}(500, 220) 
  \put(0, 0){\includegraphics[width=1.0\linewidth]{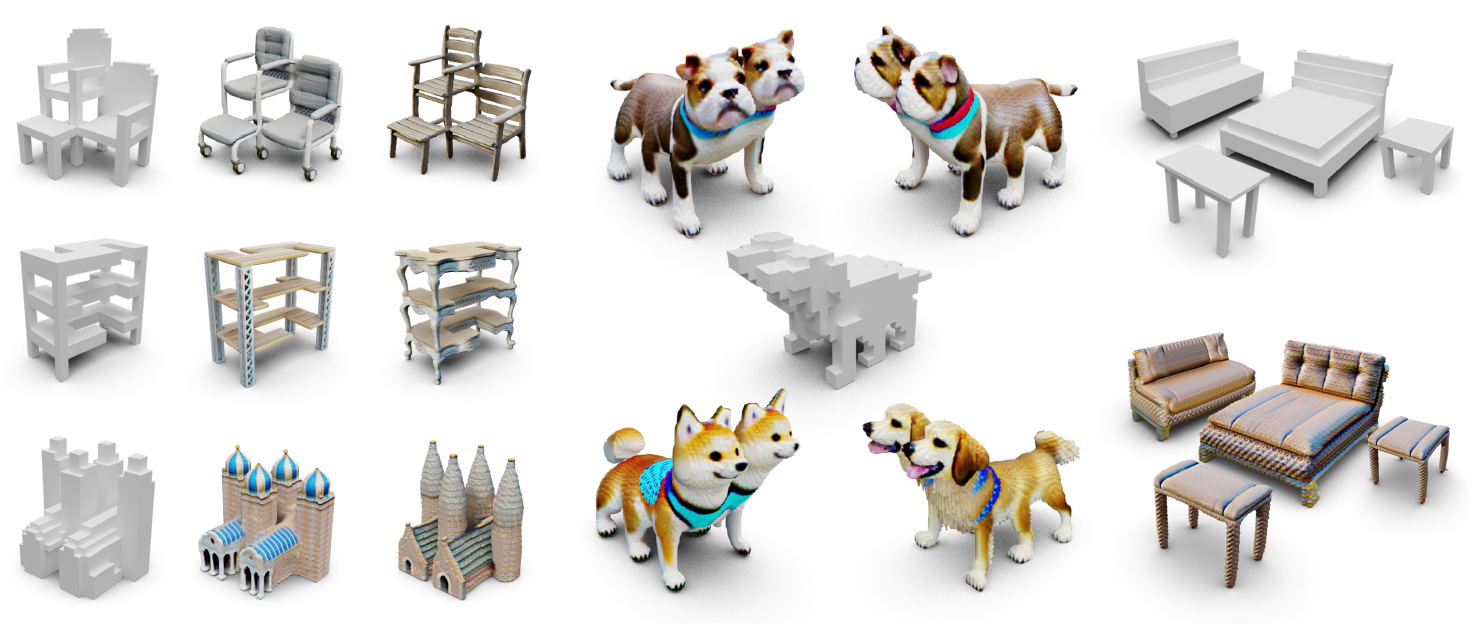}}
  \put(55, 218){\footnotesize \textit{``an office chair with}}  
  \put(50, 211){\footnotesize \textit{wheels and thick padding''}}
  \put(132, 218){\footnotesize \textit{``a rustic wooden chair}}  
  \put(132, 211){\footnotesize \textit{with rough texture look''}}
  \put(265, 214){\footnotesize \textit{``a cute bulldog''}} 
  \put(48, 144){\footnotesize \textit{``an industrial table with}}  
  \put(44, 137){\footnotesize \textit{a metal frame and wood top''}}
  \put(132, 144){\footnotesize \textit{``a Queen Anne table with}}  
  \put(134, 137){\footnotesize \textit{smooth and muted finish''}}
  \put(58, 73){\footnotesize \textit{``an Italian basilica}}     
  \put(61, 66){\footnotesize \textit{with domed roof''}}
  \put(122, 73){\footnotesize \textit{``a castle with cone-shaped}}  
  \put(122, 66){\footnotesize \textit{tower roofs and brick walls''}}
  \put(225, 97){\footnotesize \textit{``a cute}}  
  \put(222, 90){\footnotesize \textit{Shiba Inu''}}
  \put(335, 97){\footnotesize \textit{``a cute}}  
  \put(325, 90){\footnotesize \textit{golden retriever''}}
  \put(397, 115){\footnotesize \textit{``a stylish furniture piece with leather}}  
  \put(397, 108){\footnotesize \textit{and a plush surface in a natural tone''}}
  \end{picture}
  \vspace{-4mm}
  \caption{
Our 3D {\em detailizer\/} is trained using a text prompt, which defines the shape class and guides the stylization and detailization of any number of coarse 3D shapes with varied structures. Once trained, our detailizer can instantaneously (in $<$1$s$) transform a coarse proxy into a detailed 3D shape, whose overall structure respects the input proxy and the appearance and style of the generated details follows the prompt. We show results for both human-made (left) and organic (middle) shapes, with clearly out-of-distribution structures (e.g., the multi-seat ``chair'', ``SIG"-letter shaped shelvings, and two-headed dogs with six legs). On the right, when the training prompt references a generic term such as ``furniture,'' which encompasses multiple object categories with diverse structures, such as chairs, beds, stools, etc., the 3D detailizer can be reused, as a feed-forward model without retraining, to produce a collection of detailized 3D models spanning all of these categories with structural variations. These models can then be arranged to form a style-consistent 3D scene.}
  \label{fig:teaser}
\end{teaserfigure}
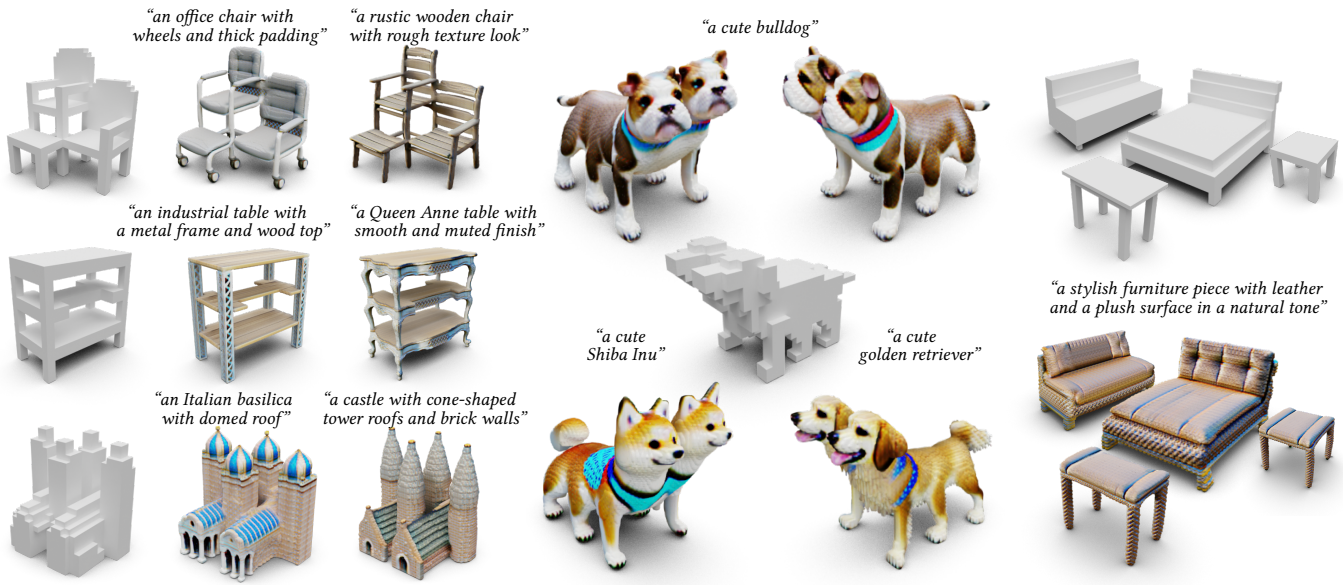



\begin{abstract}
We introduce a {\em 3D detailizer\/}, a neural model which can {\em instantaneously\/} (in <1s) transform a coarse 3D shape proxy into a high-quality asset with detailed geometry and texture as guided by an input text prompt. Our model is trained using the text prompt, which defines the shape class and characterizes the appearance and fine-grained style of the generated details. The coarse 3D proxy, which can be easily varied and adjusted (e.g., via user editing), provides structure control over the final shape.
Importantly, our detailizer is not optimized for a single shape; it is the result of {\em distilling a generative
model\/}, so that it can be reused, without retraining, to generate any number of shapes,
with varied structures, whose local details all share a consistent style and appearance.
Our detailizer training utilizes a pretrained multi-view image diffusion model, with text conditioning, to distill the foundational knowledge therein into our detailizer via Score Distillation Sampling (SDS).
To improve SDS and enable our detailizer architecture to learn generalizable features over complex structures, we train our model in two training stages to generate shapes with increasing structural complexity.
Through extensive experiments, we show that our method generates shapes of superior quality and details compared to existing text-to-3D models under varied structure control.
Our detailizer can refine a coarse shape in less than a second, making it possible to interactively author and adjust 3D shapes. 
Furthermore, the user-imposed structure control can lead to creative, and hence out-of-distribution, 3D asset generations that are beyond the current capabilities of leading text-to-3D generative models.
We demonstrate an interactive 3D modeling workflow our method enables, and its strong generalizability over styles, structures, and object categories.

\end{abstract}

%
%


%
%


\maketitle

\section{Introduction}
\label{sec:intro}

3D generative models are becoming increasingly powerful, enabling the creation of 3D content with ease from texts~\cite{poole2022dreamfusion,shi2023mvdream,li2023instant3d} or images~\cite{liu2023zero123,hong2023lrm,xiang2024structuredLatent}. However, artistic creation involves realizing the design vision of the artist, which often requires precise control over both the coarse structure and the local details of the generated object. Such precision cannot be fully achieved through text or image inputs alone.
In addition, the artist's creative exploration of the design space is greatly facilitated by the ability to quickly generate detailed 3D assets via structural variations.

\begin{figure}[!tb]
\begin{picture}(244, 147)
  \put(0, 0){\includegraphics[width=0.99\linewidth]{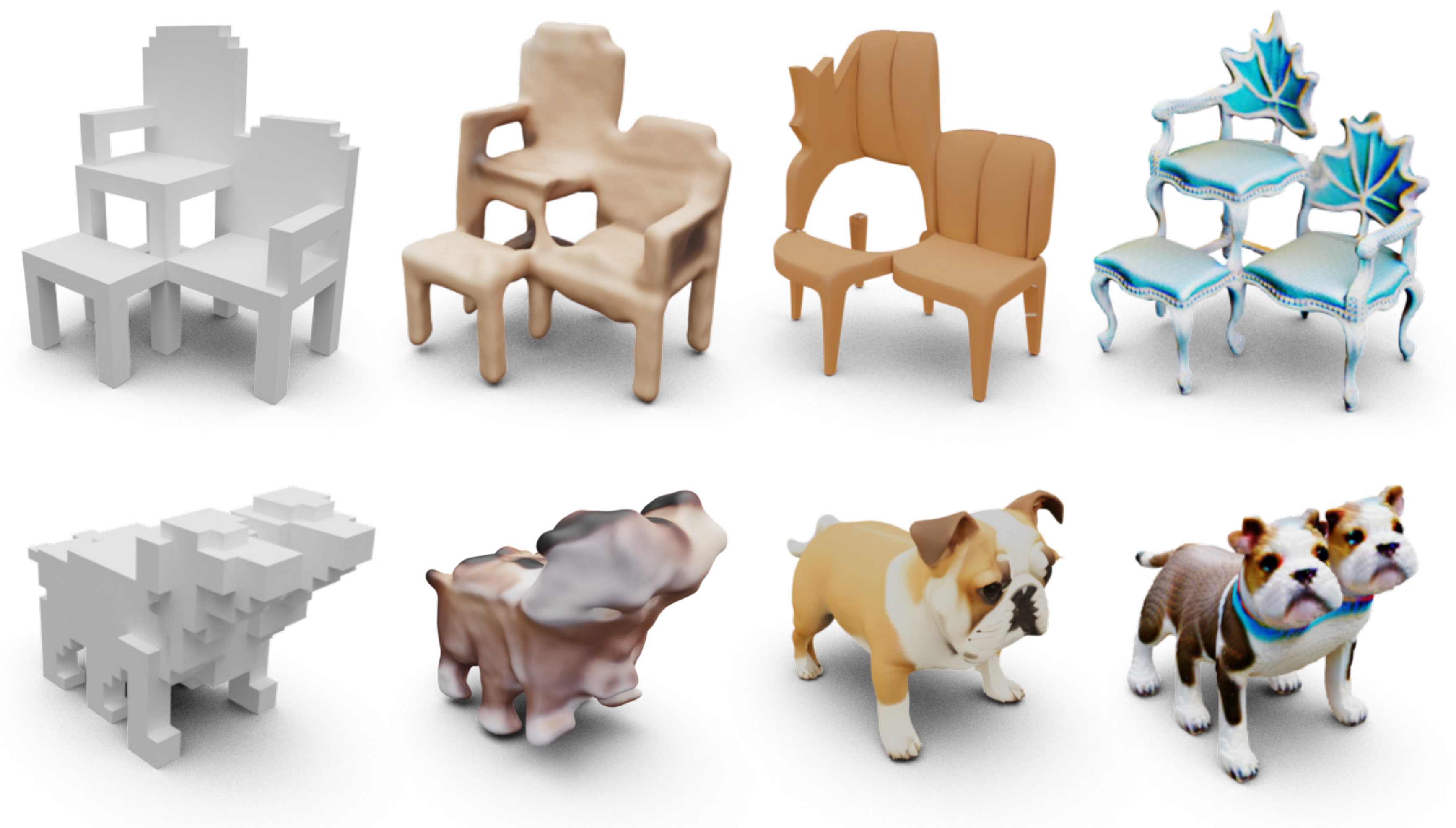}}
  \put(5, -3){\small (a) Coarse shape}
  \put(75, -3){\small (b) Coin3D}
  \put(137, -3){\small (c) CLAY}
  \put(194, -3){\small (d) Ours}
  
  \put(35, 137){\small \textit{``a chair with a backrest shaped like a large maple leaf''}}
  \put(100, 60){\small \textit{``a cute bulldog''}}
\end{picture}
\vspace{-4mm}
  \caption{Comparing to state-of-the-art generators on out-of-distribution coarse structures (a). Coin3D~\cite{dong2024coin3d} (b) is unable to produce proper textures based on the prompts, while CLAY~\cite{zhang2024clay} (c) falls short in respecting the input structures.
  Our method performs better on both fronts, e.g., maple leaf-shaped backs with curved armrests and legs as matching styles, and the multiple legs and heads of the bulldog.}
  \vspace{-1mm}
  \label{fig:related_work_fig}
\end{figure}
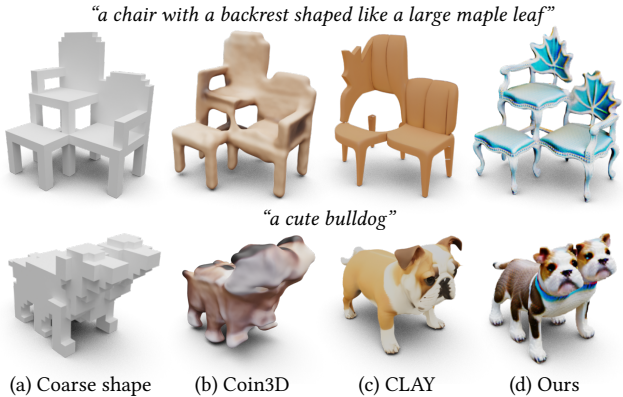

Prior works propose generative models which rely on user-defined coarse shapes to control the structure of the generated shapes, e.g., by formulating the problem as that of 3D voxel up-resolution~\cite{chen2021decorgan,chen2025decollage,chen2024shadder,shen2021dmtet,ren2024xcube}, while some others~\cite{zhang2024clay,hui2024make} develop 3D generative models with coarse shapes as conditioning.
However, as their models were trained on limited 3D shapes, they face generalizability issues and cannot produce local details of arbitrary desired styles.
Some approaches~\cite{chen2023fantasia3d,metzer2023latent} optimize the initial coarse shape via Score Distillation Sampling (SDS) using text-to-image diffusion models, which possess much stronger generalizability as they were trained on large image collections.
Yet the structures of their generated shapes often deviate from those of the input shapes. In addition, each shape takes a significant amount of time to optimize, ranging from minutes to hours.
More recently, some approaches~\cite{YunChun24Text-guidedMeshRefinement,dong2024coin3d} have improved the structure adherence by adopting finetuned multi-view diffusion models conditioned on coarse shape inputs.

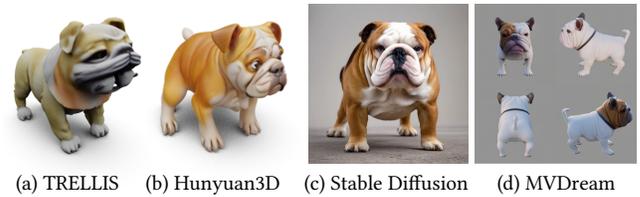
\begin{figure}[!tb]
\begin{picture}(244, 74)
  \put(0, 0){\includegraphics[width=0.99\linewidth]{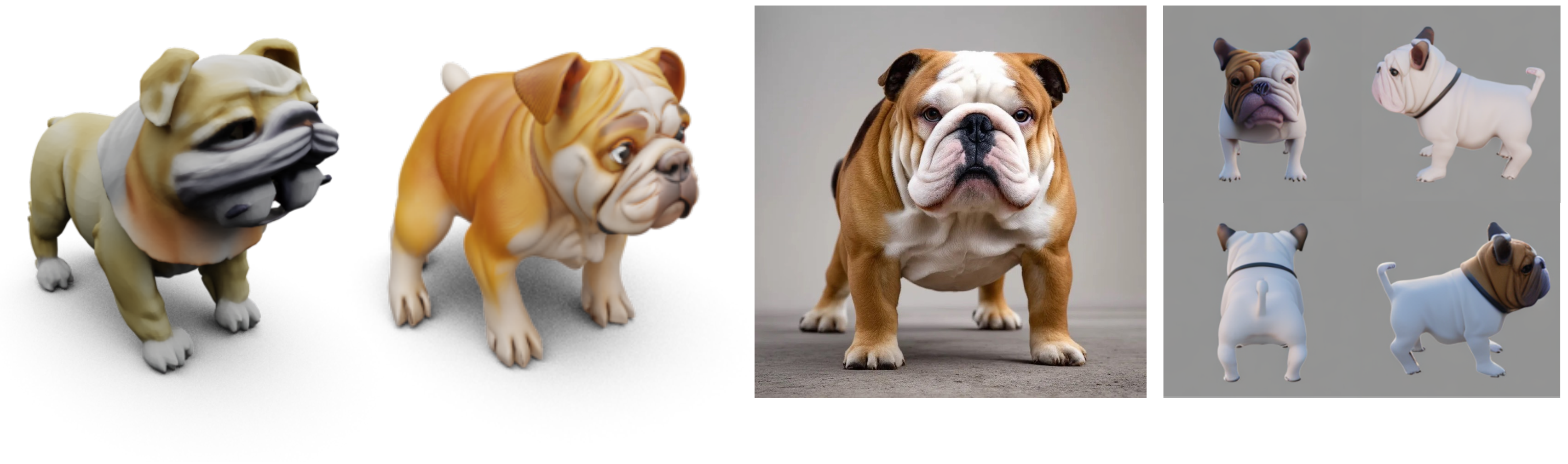}}
  \put(5, 0){\small (a) TRELLIS}
  \put(54, 0){\small (b) Hunyuan3D}
  \put(114, 0){\small (c) Stable Diffusion}
  \put(187, 0){\small (d) MVDream}
\end{picture}
\vspace{-4mm}
  \caption{State-of-the-art text-to-3D and text-to-image models are unable to generate \textbf{\textit{``a cute bulldog with two heads and six legs"}} from text.}
  \vspace{-3mm}
  \label{fig:related_work_fig_2}
\end{figure}

Despite such advances, a common issue with all methods relying on image diffusion models is that they can only generate shapes that are ``ordinary'' with respect to the pretrained diffusion models --- they typically fail when the structure of the conditioning coarse shape is out of distribution, e.g., see examples from Figures~\ref{fig:teaser} and a comparison to current generation models in Figure~\ref{fig:related_work_fig}. Figure~\ref{fig:related_work_fig_2} further shows that state-of-the-art text-to-3D and text-to-image models are unable to deal with such out-of-distribution examples.
Also, as these models perform {\em per-shape\/} optimization with no style consistency across different shapes, they cannot generate a coherent {\em collection\/} of 3D shapes with varied structures under the same style prompt, as shown by the ``furniture'' example in Figure~\ref{fig:teaser}-right.

In this work, we aim to tackle the above issues by introducing a 3D detailization model, or a {\em detailizer\/}, which is trained using a text prompt and can transform a coarse 3D shape proxy into a high-quality asset with
detailed geometry and texture. The text prompt defines the shape class and guides the stylization and detailization of any number of coarse 3D shapes with varied structures. On the other hand, the shape proxy, which can 
be easily adjusted (e.g., via user editing), provides structure control over the final shape.

Our method relies on a pretrained multi-view image diffusion model with text conditioning, to achieve the generation of different possible styles. Given a text description of the desired style, instead of optimizing a single shape with SDS as in prior works~\cite{chen2023fantasia3d,metzer2023latent,YunChun24Text-guidedMeshRefinement,dong2024coin3d}, 
we {\em distill a 3D generative model\/} as our detailizer, so that it can be reused, without retraining, to generate any number of shapes, with varied structures, whose local details all share a consistent style.
We leverage a 3D convolutional neural network (CNN) as the backbone for our detailizer, so that the locality of the convolution operators helps our model learn localized features and enables it to handle coarse shapes of arbitrary structures.
During training, we utilize a structure-matching loss defined on the rendered images of the generated shape and the input coarse shape, so the generated shape matches the structure of the input. To facilitate the learning of generating complex structures, we train the detailizer in two training stages to generate shapes with increasing structural complexity.
When the training is done for the given text prompt, our detailizer is deployable as a feed-forward model that can detailize a coarse 3D shape in {\em less than a second\/}.
This allows interactive exploration of structurally varying 3D shape designs in a common style space.

Our method is coined ART-DECO: \underline{Ar}bitrary \underline{T}ext guidance for 3D \underline{De}tailizer \underline{Co}nstruction.
We demonstrate both quantitatively and qualitatively that ART-DECO generates 3D shapes of superior 
quality and details compared to state-of-the-art 3D generative models with structure control, such as ShaDDR~\cite{chen2024shadder}, CLAY~\cite{zhang2024clay}, and Coin3D~\cite{dong2024coin3d},
especially when the input structure may be out-of-distribution and exhibit creativity; see Figure~\ref{fig:related_work_fig}.
We also show that ART-DECO can be trained using a single prompt to reference a {\em generic term\/} such as ``furniture.'' Then, the same detailizer can be reused to quickly generate structurally varying shapes spanning {\em multiple furniture categories\/} such as chairs, tables, beds, etc., which share a common style, e.g., ``leather, plush surface'', as also prescribed in the training prompt; see Figure ~\ref{fig:teaser}.
We further demonstrate an interactive 3D modeling workflow ART-DECO enables, and its generalizability to different out-of-distribution structures in an interactive application.

\section{Related work}
\label{sec:related}

\begin{figure*}
\begin{picture}(510, 200)
\centering
  \put(0, 0){\includegraphics[width=1.0\linewidth]{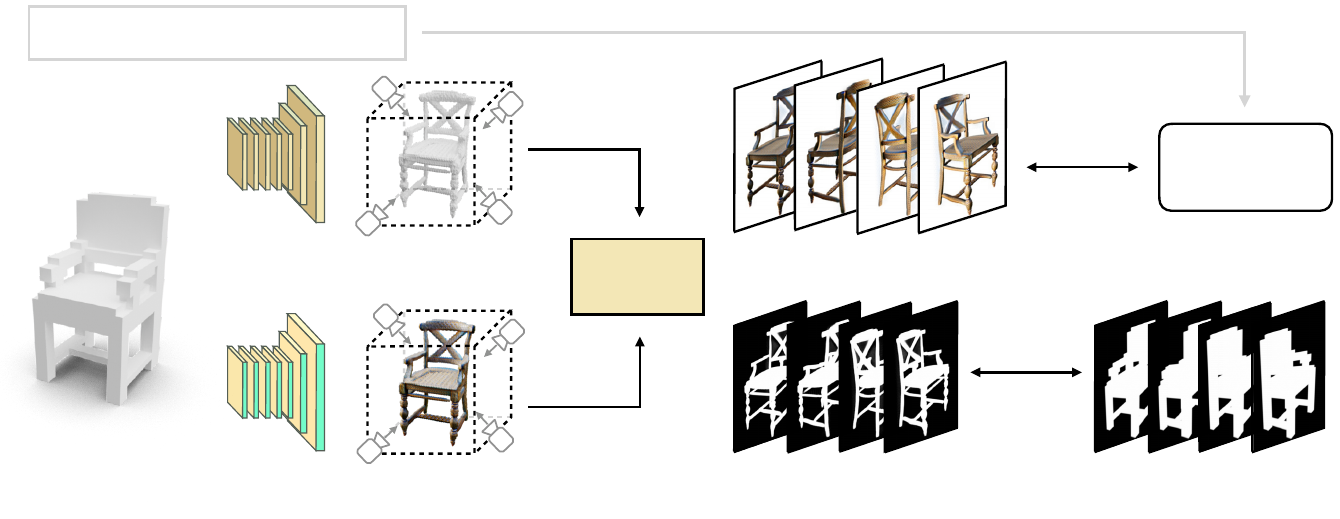}}

  \put(0, 8){\small Input coarse voxel grid}
  \put(12, 186){\textit{Text prompt: ``a farmhouse chair with a}}
  \put(12, 177){\textit{cross-back and a natural wood finish''}}

  \put(86, 97){\small Upsampling}
  \put(88.5, 88){\small networks}

  \put(222.5, 94){Volumetric}
  \put(224.5, 84){rendering}

  \put(145, 93){\small Density field}
  \put(145, 8){\small Albedo field}

  \put(402, 140){$\mathcal{L}_{SDS}$}
  \put(382, 63){$\mathcal{L}_{reg}$}
  \put(295, 8){\small Rendered mask}
  \put(435, 8){\small Coarse mask}

  \put(303, 93){\small Rendered RGB}

  \put(418, 173){\small Text condition}
  \put(454, 140){Pretrained}
  \put(454, 130.5){multi-view}
  \put(445, 121){diffusion model}
  
\end{picture}
\vspace{-7mm}
  \caption{
  Overview of the training of our detailizer. Given a coarse voxel grid and a text prompt that describes a style, two 3D convolutional networks upsample the coarse voxels into high-resolution density and albedo fields, respectively. Multi-view images are then rendered from the density and albedo fields, and a pretrained multi-view diffusion model conditioned on the text prompt is used as a prior for Score Distillation Sampling ($\mathcal{L}_{SDS}$). The regularization loss ($\mathcal{L}_{reg}$) enforces structural consistency by measuring the similarity between the masks rendered from the generated shape and those from the input coarse voxel grid. During inference time, the detailizer upsamples a coarse voxel grid into high-resolution fields in a single feedforward pass, without SDS optimization.
  }
  \vspace{-1mm}
  \label{fig:pipeline}
\end{figure*}

\paragraph{3D generative models.} 
3D generation methods went through drastic development in the recent decade. Early works mainly focus on building generative models under various 3D representations, such as voxels~\cite{choy20163d-r2n2,wu2016voxelGan}, point clouds~\cite{achlioptas2018learning,fan2017point,nichol2022pointe}, implicit fields~\cite{mescheder2019occupancynet,zhiqin2019sdf,park2019deepsdf}, and polygon meshes~\cite{nash2020polygen,shen2024spacemesh,siddiqui2024meshgpt}. These methods are trained with various generative frameworks such as variational autoencoders~\cite{kingma2013vae}, generative adversarial networks~\cite{goodfellow2020gan}, and denoising diffusion probabilistic models~\cite{song2020diffusion,ho2020ddpm}. Due to data scarcity, these methods have limited generalization ability beyond the training distribution.

With the success of image diffusion models~\cite{saharia2022imagen,rombach2022stablediffusion}, DreamFusion~\cite{poole2022dreamfusion} and subsequent works~\cite{wang2023scorechainjacobian,lin2023magic3d,melas2023realfusion, chen2023fantasia3d} distill the 2D image prior in the diffusion models into 3D shapes by optimizing individual shapes with Score Distillation Sampling (SDS), therefore achieving zero-shot text-to-3D generation.
Several works~\cite{xie2024latte3d,lorraine2023att3d,qian2024atom} also attempt to distill feed-forward text-to-3D generative models with SDS.
While their generalization ability has been greatly improved, they still lack 3D understanding, which often leads to suboptimal results such as the multi-face Janus problem~\cite{chen2023fantasia3d}. Zero123~\cite{liu2023zero123} finetunes image diffusion models on large 3D object datasets such as Objaverse~\cite{deitke2023objaverse}, to make the diffusion model capable of generating novel-view images conditioned on a single input image and the target viewpoints. Later works including MVDream~\cite{shi2023mvdream} and others~\cite{shi2023zero123++,long2024wonder3d,liu2023syncdreamer,xiang2023image} focus on improving multi-view consistency by generating all the views together and introducing extra information to condition the diffusion process. These multi-view images can then be used to reconstruct a 3D shape via differentiable rendering~\cite{wang2021neus,mildenhall2021nerf,kerbl2023gaussian} or via a feed-forward 3D shape reconstruction network~\cite{hong2023lrm,li2023instant3d}. 

Recently, a few methods~\cite{xiang2024structuredLatent,zhang2024clay,chen20243dtopia,ren2024xcube,wu2024direct3d,li2024craftsman,hui2024make} have been proposed to train 3D diffusion models directly from large 3D datasets, bypassing the intermediate image diffusion model. Nevertheless, the existing 3D generative models primarily focus on text or image-conditioned generation. They often lack precise control over the overall structure of the generated 3D shapes, therefore making it difficult to be incorporated into an artist’s workflow. In contrast, our method trains a feed-forward model that detailizes a coarse shape with a style specified by a text prompt, making it possible to create 3D shapes in an interactive manner where an artist can make adjustments to the coarse shape for refinement.

\paragraph{Geometric detailization.}
Many traditional~\cite{kajiya1989fur,neyret1998volumetric, zhou2006mesh} and learning-based~\cite{hertz2020deeptexturesynthesis,berkiten2017learning,yifan2022geometryconsistent} approaches have been proposed to synthesize geometric details on coarse shapes by transferring geometric textures, which are often represented as displacement maps or geometric texture patches, from detailed shapes. In addition, neural subdivision~\cite{Liu:Subdivision:2020} and subsequent works~\cite{shen2021dmtet,chen2021neuralmc,chen2023neuralprogressivemesh} can learn local geometric details in training shapes and apply them to new shapes. In another line of work, methods that are based on voxel representation can generate geometric details by replicating local voxel patches in detailed reference shapes~\cite{chen2021decorgan,chen2025decollage,chen2024shadder,Sun22patchRD}. However, these methods all require detailed 3D exemplar shapes provided as style references, which limits their capability when the exemplar shapes with desired style are scarce or unavailable. To mitigate the issue of data scarcity, some recent methods~\cite{chen2023fantasia3d,metzer2023latent,Gao23TextDeformer,michel2022text2mesh} propose to utilize image-space supervision provided by CLIP~\cite{radford2021clip} or SDS, where the style is provided by an input text. However, these methods take a long time to converge, hindering the possibility of an interactive modeling experience. Most recent works~\cite{YunChun24Text-guidedMeshRefinement,dong2024coin3d,Liu_2024_CVPR_sherpa3d} leverage multi-view image diffusion models conditioned on input coarse shapes, which provide faster convergence and better structure adherence. Despite these advancements, a common issue with these methods is that they can only generate shapes within the training distribution and fail when the structure of the conditioning coarse shape is uncommon. Our work addresses this issue by distilling a generative model which is designed to be generalizable and trained with increasing structure complexity.

\section{Method}
\label{sec:method}

In this section, we detail the design and the training of our detailizer. An overview is provided in Figure ~\ref{fig:pipeline}. Our detailizer upsamples an input coarse shape represented as binary occupancy voxels into a detailed 3D shape represented by a volume radiance field. Specifically, given an input coarse voxel grid of resolution $k^3$, our detailizer networks will generate two voxel grids of resolution $K^3$ to store a density field and a albedo field. The generated shape can then be rendered via volumetric rendering from the two fields; alternatively, a mesh with textures can be exported to be visualized. In this paper, we always export meshes for visualization. We use $k=32$ and $K=128$ in our experiments.

\subsection{Network architecture}
\label{sec:method_net}

We utilize 3D convolutional networks as backbone for our detailizer, as inspired by prior voxel detailization works~\cite{chen2021decorgan,chen2025decollage,chen2024shadder} which showed great generalizability on arbitrary input coarse shapes.
The detailed network architecture can be found in the supplementary.
The input to our detailizer is a coarse occupancy voxel grid $v \in \{0,1\}^{k\times k\times k}$. The detailizer has two separate upsampling networks: $\mathbf{G}_{d}$ for generating the density field $v'_d = \mathbf{G}_{d}(v) \in [0,+\infty)^{K\times K\times K}$, and $\mathbf{G}_{a}$ for the albedo field $v_a = \mathbf{G}_{a}(v) \in [0,1]^{K\times K\times K\times 3}$, where each voxel stores an RGB color.
To enforce that the generated shape follows the structure of the input voxels, we dilate the input voxel grid $v$ by one voxel~\cite{chen2024shadder} and then upsample it to $K^3$ resolution via nearest neighbor to obtain a binary mask $v_m$. We then apply the mask to the output density field $v'_d$ via element-wise multiplication to obtain the final density field: $v_d = v'_d \cdot v_m$. This masking step prevents the network from generating artifacts or unnecessary details far away from the structure of the input coarse shape, and allows the network to focus its capacity on generating plausible details only within the valid region $v_m$.

\subsection{Training}
\label{subsec:training}

To train the detailizer, we run the networks on a training set of coarse voxel grids. For each voxel grid, we generate the density field $v_d$ and the albedo field $v_a$, render multi-view images from them via volumetric rendering~\cite{mildenhall2021nerf}, and leverage a pretrained multi-view diffusion model to provide training supervision via SDS~\cite{poole2022dreamfusion,shi2023mvdream}.
Formally, given the density field $v_d$, the albedo field $v_a$, the camera parameters $\mathbf{c}_i$ (for the i-th view), the volumetric rendering function $\mathbf{R}(\cdot)$, a text prompt $y$, and the pretrained multi-view diffusion model $\epsilon_{\theta}$, we first render multi-view images $\mathbf{x}_i=\mathbf{R}(v_d, v_a; \mathbf{c}_i)$.
Note that $\epsilon_{\theta}$ operates on multi-view images, so in each iteration, we render a set of images $\{\mathbf{x}_i\}_{i=1,...,N}$ from the same shape, and then we can use $\epsilon_{\theta}$ to obtain the denoised images of $\mathbf{x}_i$: $\{\hat{\mathbf{x}}_i\}=\epsilon_{\theta}(\{\mathbf{x}_i\}_t; y,\{\mathbf{c}_i\}, t)$, where $t$ is the time step and $\{\mathbf{x}_i\}_t$ are input images after adding noise $\epsilon$ for time step $t$. Now we can define our multi-view SDS loss as
\begin{align}
    \mathcal{L}_{SDS} = \mathbb{E}_{t, \epsilon, \mathbf{c}_i} \| \mathbf{x}_i-\hat{\mathbf{x}}_i \|_{2}^{2}.
\end{align}
We use MVDream~\cite{shi2023mvdream} as our pretrained multi-view diffusion model, which is a model fine-tuned from Stable Diffusion~\cite{rombach2022stablediffusion} on the Objaverse dataset~\cite{deitke2023objaverse} to generate $N=4$ views at the same time.

SDS alone does not guarantee that the structure of the generated shape is consistent with the input coarse voxels. Even with the masking step described in Section ~\ref{sec:method_net} so that the network can only generate details in the coarse voxels' vicinity, the network often produces 3D shapes that miss certain structures, such as armrests in a chair. Therefore, to ensure the generated shape fully respects the structure of the input coarse voxels, we enforce the rendered mask (the alpha/transparency channel in the rendered image) of the generated shape to be similar to the rendered mask of the input coarse voxel grid when the masks are rendered from the same camera pose. We formulate this constraint as a regularization loss
\begin{align}
    \mathcal{L}_{reg} = \mathbb{E}_{\mathbf{c}_i}  \| \mathbf{m}_i-\hat{\mathbf{m}}_i \|_{2}^{2}
\end{align}
where $\mathbf{m}_i$ is the rendered mask of the generated shape from camera pose $\mathbf{c}_i$, and $\hat{\mathbf{m}}_i$ is the rendered mask from the input coarse voxels.

The final loss function to train our model is defined as
\begin{align}
    \mathcal{L} = \mathcal{L}_{SDS} + \lambda_{reg}\cdot\mathcal{L}_{reg}
\end{align}
where $\lambda_{reg}$ is a hyper-parameter balancing between distilling plausible shapes from SDS and preserving the structure of the inputs. During training, we set $\lambda_{reg}=10^{4}$, and gradually reduce the value to $10$, allowing the network to focus on generating structures in the early stages of training and refining local details later.

\subsection{Data}

Our input text prompt specifies the desired style and indicates the generic shape category of the generated shapes, such as chairs; however, it does not provide detailed structural descriptions.
During training, we need coarse voxel grids of that shape category to serve as input to our detailizer.
For a specific shape category, if a decently large 3D dataset is available, we simply voxelize the existing shapes into coarse voxel grids as our training set.
Otherwise, we download a few shapes (16-32) from the Internet, an adopt a similar strategy to that of DECOLLAGE~\cite{chen2025decollage} to generate a large number of coarse voxel grids with diverse structures via data augmentation. Specifically, we randomly scale the existing shapes in $x$, $y$, and $z$ directions, and apply random rotations and merge multiple shapes into a single shape to enrich the geometric variation. See Section ~\ref{sec:exps} for more details about the shapes used in our experiments.

\begin{figure}
\begin{picture}(244, 135)
  \put(50, 0){\includegraphics[width=0.8\linewidth]{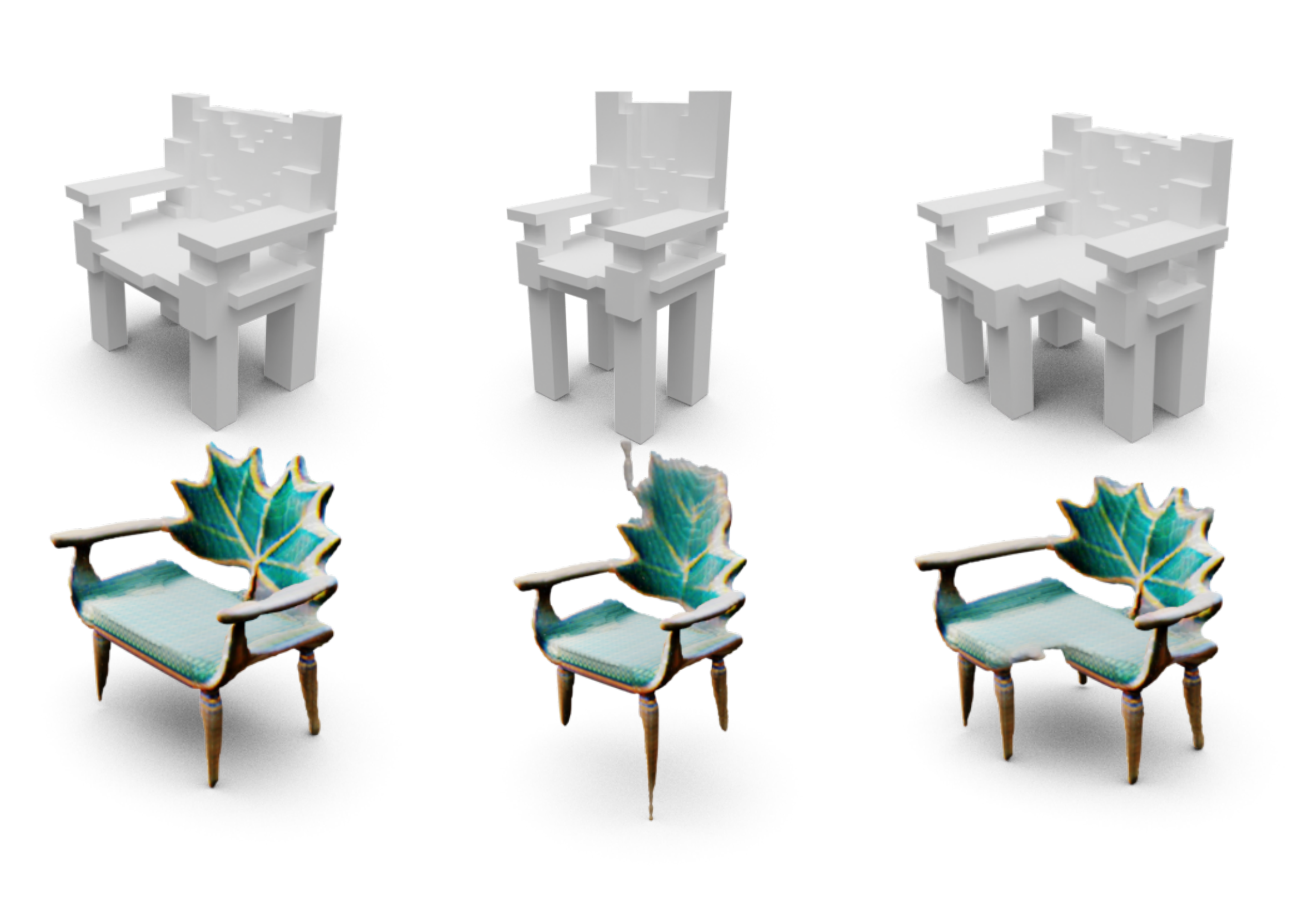}}



  \put(13, 100){\small Input}
  \put(0, 90){\small coarse voxels}

  \put(10, 45){\small Output}
  \put(11, 35){\small shapes}

  \put(66, 63){\small (a)}
  \put(128, 63){\small (b)}
  \put(197, 63){\small (c)}

  \put(60, 125){\small \textit{``a chair with a backrest shaped like a large maple leaf''}}

\end{picture}
\vspace{-13mm}
  \caption{Strong generalizability of 3D convolutional networks. During the first training stage, our detailizer is trained on a single coarse voxel grid to generate a single detailed shape (a). After training, when tested on different structures, our detailizer demonstrates strong generalizability and produces reasonable results (b, c). These results are not perfect, but they can be good initial points for the image diffusion model to refine in our second training stage, where we use multiple coarse shapes.}
  \vspace{-5mm}
  \label{fig:conv_generalizability}
\end{figure}
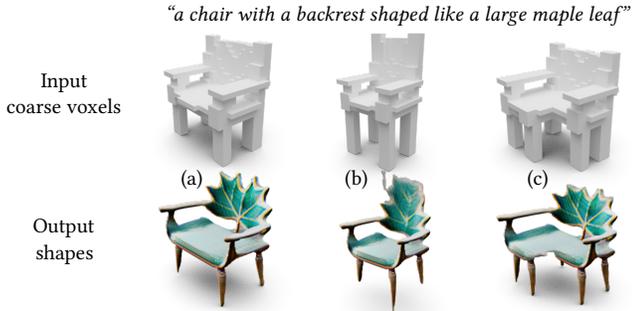

\subsection{Two-stage Learning}
\label{sec:method_curriculum}

When our detailizer was trained with all the coarse shapes in the training set (which contains both simple and complex structures; see Figure~\ref{fig:simple_complex_coarse} for examples), it failed to generate certain structures, as shown in Figure~\ref{fig:ablation_initialization}.
This is because the multi-view image diffusion model tends to prioritize generating shapes that align with its learned biases, i.e., shapes with simple and common structures. Therefore, when the model is trained with complex structures from the very beginning, it learns to simplify the structure rather than adhering to it.
To address this, we have observed that 3D convolutional networks have strong generalizability even when trained on a single simple shape, thanks to the locality and inductive bias of convolution operations.
As shown in Figure~\ref{fig:conv_generalizability}, our detailizer trained on a single coarse voxel grid generalizes reasonably well on other inputs during inference, even when the input has complex structures.
This model with single-shape training can be a good initialization for multi-shape training by providing good initial shapes for the image diffusion model to refine.
Therefore, we employ a two-stage training scheme.
In the first stage, we train the model on a single input coarse voxel grid with a simple structure.
And in the second stage, we keep training the model on all the coarse shapes in our training set until convergence.

In our experiments, we train individual models for different text prompts. The first stage of training takes about 1.5 hours and the second stage takes about 3.5 hours on a single NVIDIA 4090 GPU.

\section{Experiments}
\label{sec:exps}

In this section, we test our method on text prompts describing different styles from various shape categories. We show that our detailizer can generate high-quality shapes that adhere to both the text prompt and the structure of the input coarse shape. We compare with state-of-the-art 3D generative models with text and structural control in Section ~\ref{sec:text_guided_detailization} to demonstrate the effectiveness of our proposed approach. We then validate our design via ablation study in Section ~\ref{sec:ablation_study}. Finally, we showcase several applications enabled by our method in Section ~\ref{sec:application}, including style-consistent detailization for shapes in multiple categories, and an interactive modeling application.
\paragraph{Datasets.}
We evaluate our model on seven shape categories: chair, table, couch, bed, building, animal, and cake. For each category, we collect 1,000 to 2,000 shapes from ShapeNet~\cite{chang2015shapenet}, 3D Warehouse~\cite{3dwarehouse}, and Objaverse~\cite{deitke2023objaverse} and voxelize them to obtain coarse voxels for training. See supplementary material for details.

\paragraph{Evaluation metrics.} For quantitative evaluation, we adopt the Render-FID and CLIP Score proposed in LATTE3D~\cite{xie2024latte3d} to evaluate the fidelity of the generated shapes and their consistency against the text prompt. \textit{Render-FID} measures the Fr\'echet Inception Distance (FID) between the rendered images of the generated shapes and the images sampled from Stable Diffusion with the same text prompts. It evaluates how closely the generated shapes align with the 2D prior in the image diffusion model. \textit{CLIP Score} measures the average CLIP score~\cite{radford2021clip} between the text prompt and each rendered image of the generated shape. It evaluates how closely the generated shapes align with the input text prompt. We also adopt Strict-IoU and Loose-IoU proposed in DECOR-GAN~\cite{chen2021decorgan} to measure the consistency between the structures of the generated shape and its input coarse voxel grid. \textit{Strict-IoU} computes the Intersection over Union (IoU) between the the input voxels and the occupancy voxels obtained by voxelizing the generated shape. \textit{Loose-IoU} is a less restrictive version of Strict-IoU which computes the proportion of occupied voxels in the input voxels that are also occupied in the occupancy voxels of the generated shape. 
Generating a single shape using CLAY's commercial product, Rodin, takes about 5-10 minutes, while Coin3D requires about 25 minutes to optimize one, which makes both approaches impractical for large-scale automatic evaluation. Therefore, we randomly select 8 text prompts for testing. To prepare the voxel grids for testing, for each text prompt, we randomly sample 10 coarse shapes in the same way we obtained the training coarse shapes. We run all methods on those shapes, compute the metrics, and take the average. We report the average across all text prompts in the paper, and the average for each text prompt in the supplementary material.

\begin{table}[t]
    \begin{center}
    \caption{Quantitative comparison of text-guided 3D generation with structural control. Metrics are averaged over the evaluated text prompts.
    }
    \label{tab:comparison}
    \vspace{-3mm}
    \resizebox{1.0\columnwidth}{!}{
    \begin{tabular}{lcccc}
    \toprule
        & Render-FID $\downarrow$ & CLIP score $\uparrow$ & Strict-IoU $\uparrow$ & Loose-IoU $\uparrow$ \\
    \midrule
       ShaDDR & 188.143 & 23.329 & 0.612 & 0.706 \\
       Coin3D & 200.935 & 20.250 & 0.566 & 0.678 \\
       CLAY & 175.402 & 23.943 & 0.626 & 0.725 \\
       Ours & \textbf{171.703} & \textbf{26.047} & \textbf{0.639} & \textbf{0.739} \\
    \bottomrule
    \end{tabular}
    }
    \end{center}
    \vspace{-4mm}
\end{table}

\subsection{Text-guided detailization}
\label{sec:text_guided_detailization}

We compare our method with several methods that can generate 3D shapes with structural control: CLAY~\cite{zhang2024clay}, Coin3D~\cite{dong2024coin3d}, and ShaDDR~\cite{chen2024shadder}. 
\textit{CLAY} is a feed-forward model trained directly on large 3D datasets. It is able to take a coarse voxel grid as structural guidance in addition to its text conditioning. However, its code is not publicly available. We instead use Rodin\footnote{{\protect\url{https://hyper3d.ai/}}}, a commercial product built on and powered by CLAY, as a substitute for testing.
%
\textit{Coin3D} is an optimization-based framework for refining coarse shapes.
We run the officially released code for testing in our experiments.
\textit{ShaDDR} is a shape detailization model whose network architecture and input/output closely resemble those of our detailizer. However, ShaDDR requires ground truth 3D shapes as style reference. Therefore, for each text prompt, we use one detailed shape generated by our method for training ShaDDR. We provide the training shapes in the supplementary and verify that the shapes are in good quality.
%

We showcase qualitative results of our method in Figure ~\ref{fig:additional_results_2_1}, and compare with other methods in Figure ~\ref{fig:additional_results_1_1}.
In Figure ~\ref{fig:additional_results_1_1},
Coin3D struggles to generate accurate geometry and texture when conditioned on complex structures. This is primarily because the multi-view diffusion model that Coin3D is based on does not generalize well to uncommon structures. We provide additional results in the supplementary material to show the intermediate outputs from Coin3D and to demonstrate that Coin3D works on simple structures but fails on complex structures.
ShaDDR was designed to be trained with clean, high-quality 3D shapes. Therefore, when trained with noisy shapes obtained from SDS optimization, the quality of ShaDDR's output becomes unsatisfactory.
%
CLAY has cleaner geometry and better structure preservation as it was trained on real 3D data. Nonetheless, it often fails to generate detailed local geometries and textures. CLAY can also deviate significantly from the structure of the input coarse shape if the structure is uncommon, see the table and animal examples in Figure ~\ref{fig:additional_results_1_1}.
In comparison, our method can effectively handle both out-of-distribution coarse structures and creative text prompts. It can provide users with flexible control over the structure and the style while generating high-quality shapes.
%
%
We report quantitative comparisons in Table~\ref{tab:comparison}, which further demonstrates the superior performance of our method. We also provide a user study in the supplementary to show that participants clearly prefer our results over alternatives.

\subsection{Ablation Study}
\label{sec:ablation_study}

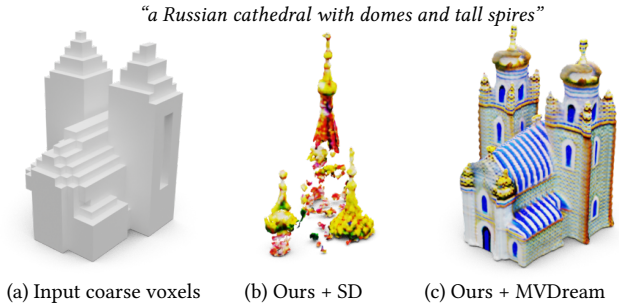
\begin{figure}
\begin{picture}(244, 120)
  \put(0, 0){\includegraphics[width=1\linewidth]{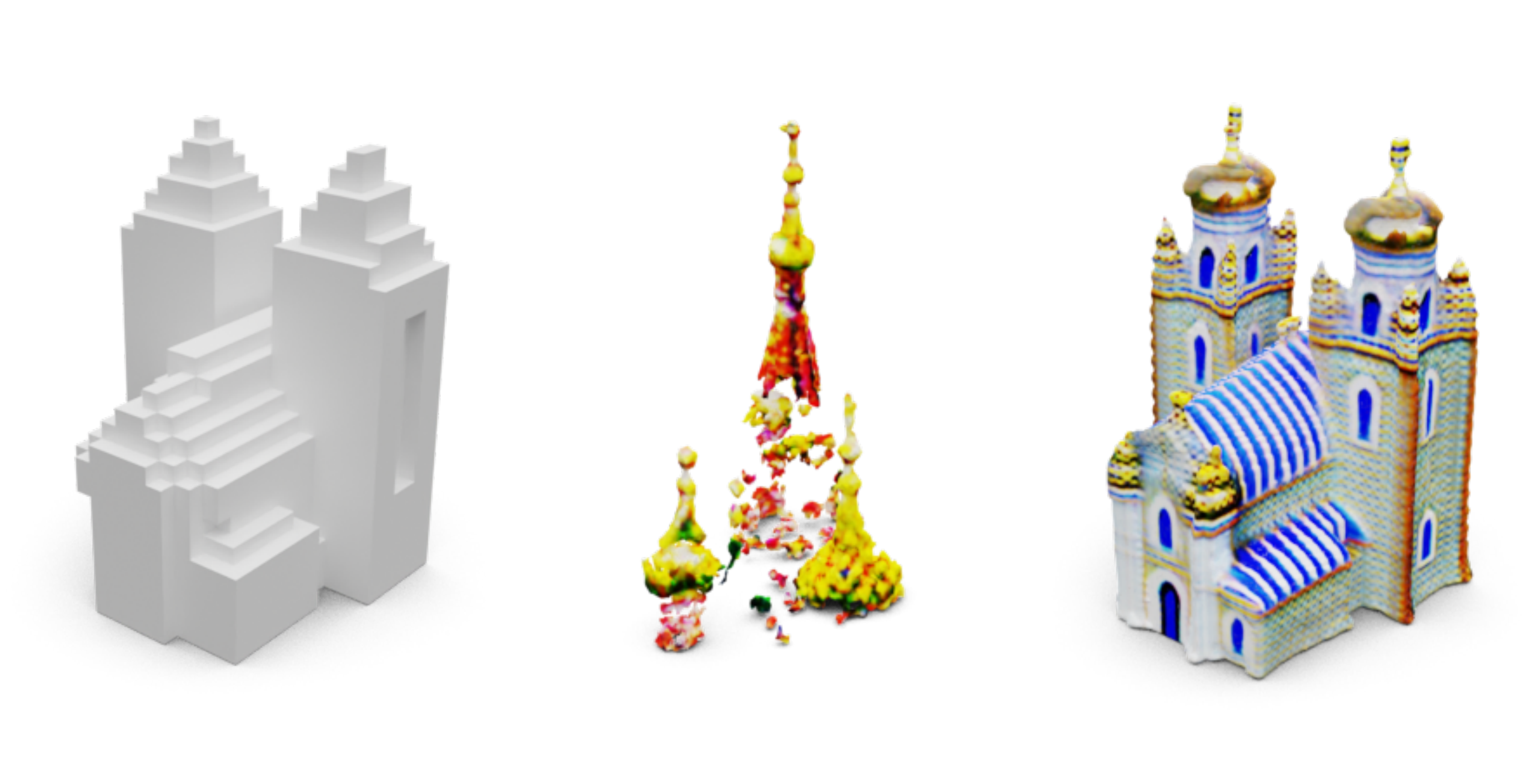}}
  \put(5, 5){\small (a) Input coarse voxels }

  \put(95, 5){\small (b) Ours + SD}

  \put(163, 5){\small (c) Ours + MVDream}

  \put(55, 110){\small \textit{``a Russian cathedral with domes and tall spires''}}

\end{picture}
 \vspace{-6mm}
  \caption{Ablation study on the choice of image diffusion models. (a) We train both models with the same coarse input in the first stage. (b) Our method with a single-view diffusion model (Stable Diffusion) as SDS guidance struggles to produce a complete shape. (c) Our method with a multi-view diffusion model (MVDream) as guidance produces high-quality results.}
  \vspace{-4mm}
  \label{fig:ablation_diffusion_model}
\end{figure}






\begin{figure}
\begin{picture}(244, 75)
  \put(0, 0){\includegraphics[width=1\linewidth]{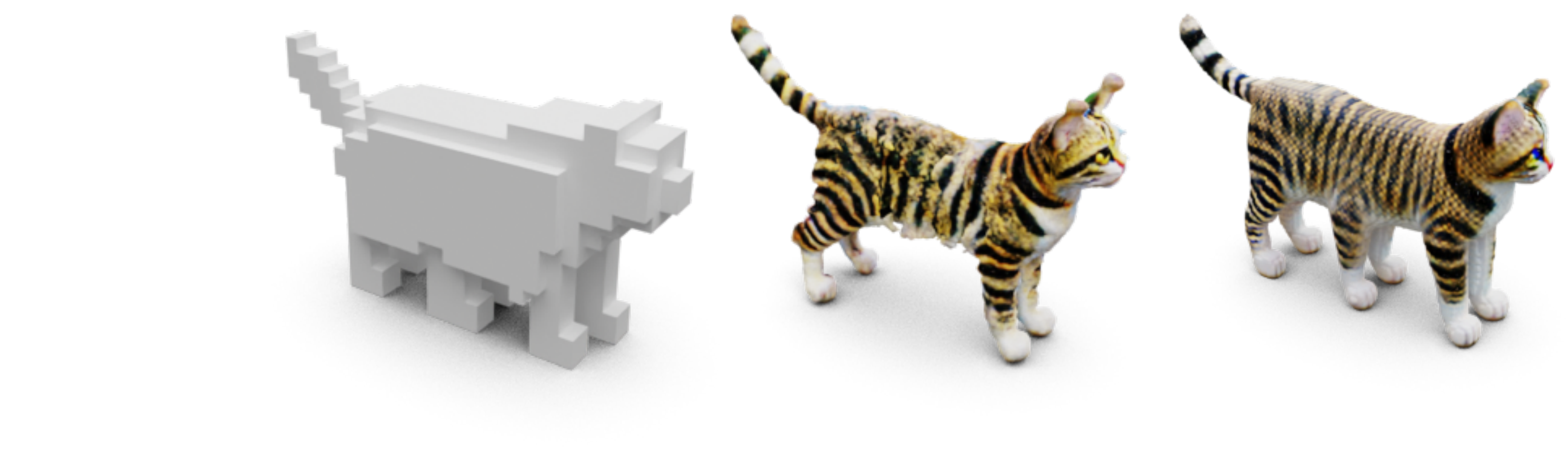}}
  \put(40, 0){\small (a) Input coarse voxels}
  \put(130, 0){\small (b) Baseline}
  \put(200, 0){\small (c) Ours}

  \put(0, 35){\small \textit{``a short-haired}}
  \put(9, 25){\small \textit{tabby cat''}}

\end{picture}
  \caption{Ablation study on the masked MVDream baseline. The baseline (b) fails to preserve the input structure, while our method (c) fully respects it.
  }
  \label{fig:baseline_comp}
  \vspace{-3mm}
\end{figure}
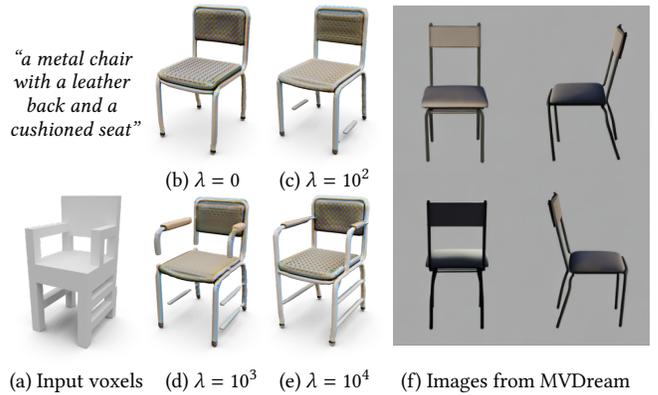
\begin{figure}
\begin{picture}(244, 152)
  \put(0, 0){\includegraphics[width=1\linewidth]{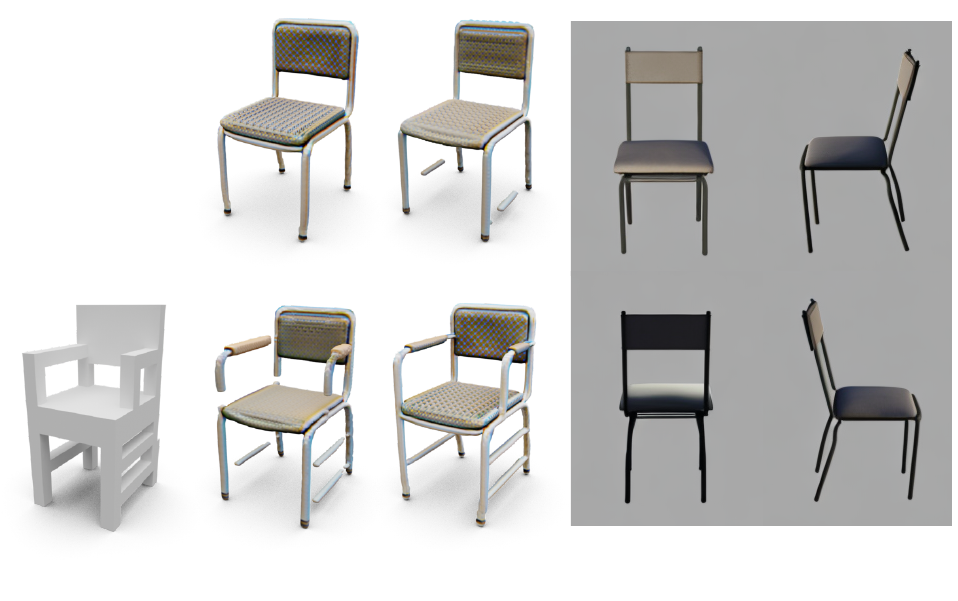}}
  \put(0, 3){\small (a) Input voxels}

  \put(1, 125){\small \textit{``a metal chair}}
  \put(1, 116){\small \textit{with a leather}}
  \put(6, 107){\small \textit{back and a}}
  \put(0, 98){\small \textit{cushioned seat''}}

  \put(59, 79){\small (b) $\lambda = 0$}
  \put(102, 79){\small (c) $\lambda = 10^2$}

  \put(59, 3){\small (d) $\lambda = 10^3$}
  \put(102, 3){\small (e) $\lambda = 10^4$}

  \put(148, 3){\small (f) Images from MVDream}
  
\end{picture}
  \caption{Ablation study on different regularization weights $\lambda_{reg}$. (b-e) show that with increasing $\lambda_{reg}$, the generated shape becomes more consistent with the input coarse structure (a). We also show example images sampled from MVDream in (f), which tends to produce simple structures.}
  \label{fig:ablation_reg_loss}
  \vspace{-5mm}
\end{figure}

\paragraph{Image diffusion model: single-view vs. multi-view.}
Since we leverage pretrained image diffusion models and distill their knowledge into our detailizer, different diffusion models can significantly affect the generation quality of 3D shapes. Therefore, we conduct an ablation study using two different types of diffusion models: Stable Diffusion, a single-view image diffusion model trained on natural images, which possesses 2D image prior; and MVDream~\cite{shi2023mvdream}, a multi-view image diffusion model finetuned on multi-view rendered images of 3D objects, thus incorporating additional 3D prior. In both settings, we use the same network and loss functions, with the only difference being the choice of image diffusion models.

Figure~\ref{fig:ablation_diffusion_model} shows the qualitative comparison of the 3D shapes generated by our method using different image diffusion models. Stable Diffusion cannot synthesize a reasonable shape while MVDream can generate a 3D shape with significantly improved geometric and texture details, which demonstrates the importance of the 3D prior in the image diffusion model and the benefit of having multi-view-consistent distillation during SDS.

\paragraph{Masked MVDream baseline.} To show the necessity of distilling a generative neural network rather than distilling a single shape, we adopt MVDream to directly optimize a neural field while enforcing structural control by masking the neural field with the input coarse voxels. As shown in Figure~\ref{fig:baseline_comp} (b), the baseline fails to preserve the input coarse structure, as MVDream's multi-view diffusion model lacks prior knowledge of atypical structures, such as a cat with six legs. In contrast, our model, although trained on coarse voxels of \textit{four-legged} animals using the same multi-view diffusion model, demonstrates strong generalization ability to unseen structures, thanks to the inductive bias of convolutional networks.

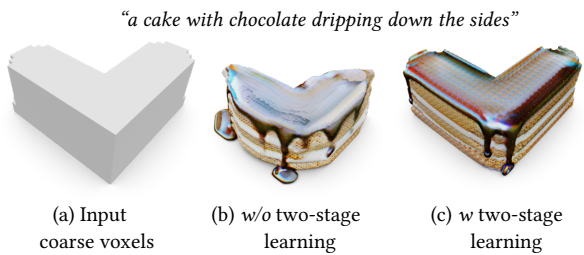
\begin{figure}
\begin{picture}(244, 95)
  \put(0, 0){\includegraphics[width=1\linewidth]{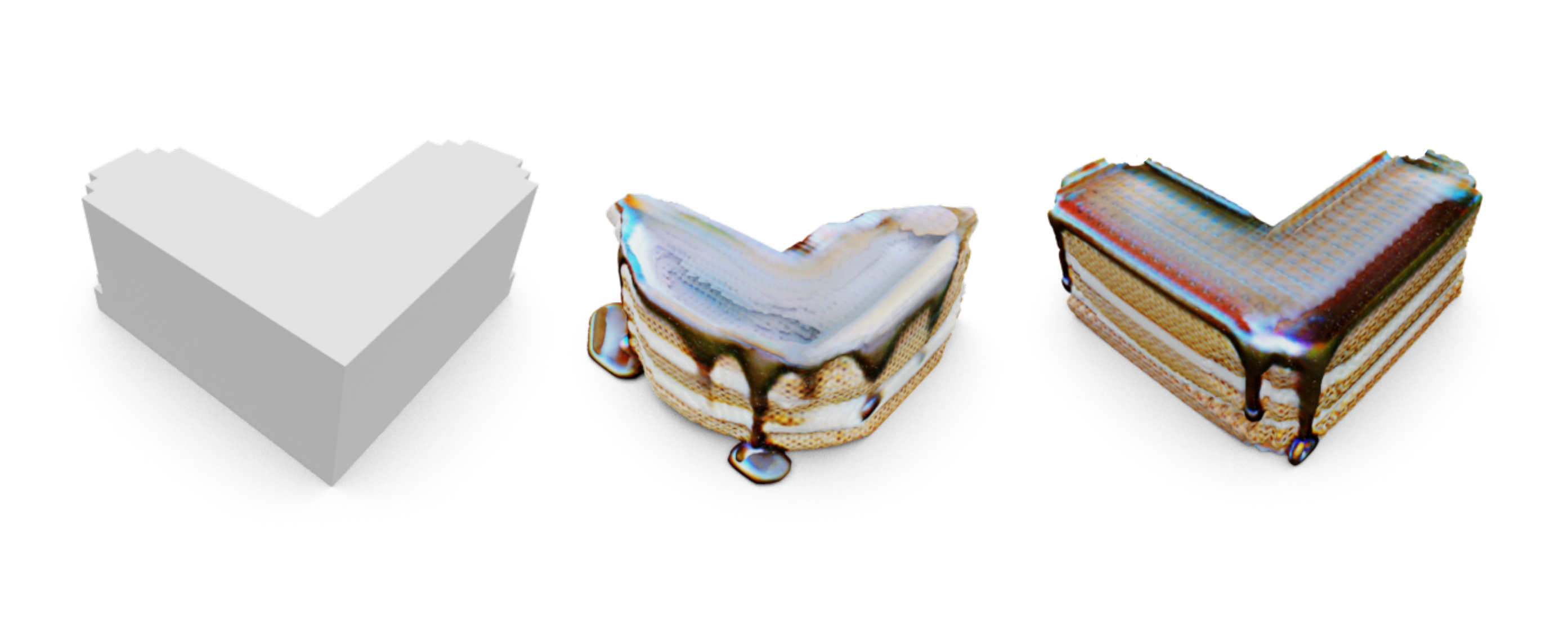}}
  \put(30, 10){\small (a) Input}
  \put(25, 0){\small coarse voxels}

  \put(90, 10){\small (b) \textit{w/o} two-stage}
  \put(110, 0){\small learning}

  \put(173, 10){\small (c) \textit{w} two-stage}
  \put(188, 0){\small learning}

  \put(55, 84){\small \textit{``a cake with chocolate dripping down the sides''}}

\end{picture}
\vspace{-6mm}
  \caption{Ablation study on two-stage learning. When skipping the first-stage training, the detailizer learns to simplify the structure of the generated shape (b) so the shape resembles a round, real-life cake; it does not strictly follow the structure of the input coarse shape (a). With both training stages, the detailizer correctly learns to generate the shape (c) in an uncommon structure. The regularization loss is used in both settings. }
  \vspace{-3mm}
  \label{fig:ablation_initialization}
\end{figure}

\begin{table}[t]
    \begin{center}
    \caption{ Quantitative ablation study on the proposed regularization loss. Metrics are averaged over the evaluated text prompts.
    }
    \label{tab:reg_ablation}
    \vspace{-3mm}
    \resizebox{1.0\columnwidth}{!}{
    \begin{tabular}{lcccc}
    \toprule
        & Render-FID $\downarrow$ & CLIP score $\uparrow$ & Strict-IoU $\uparrow$ & Loose-IoU $\uparrow$ \\
    \midrule
       $\lambda_{reg}=0$ & \textbf{171.388} & 25.718 & 0.585 & 0.693 \\
       $\lambda_{reg}=10^{2}$ & 171.684 & 25.621 & 0.584 & 0.690 \\
       $\lambda_{reg}=10^{3}$ & 171.552 & 25.423 & 0.623 & 0.728 \\
       $\lambda_{reg}=10^{4}$ & 171.703 & \textbf{26.047} & \textbf{0.639} & \textbf{0.739} \\
    \bottomrule
    \end{tabular}
    }
    \end{center}
    \vspace{-4mm}
\end{table}

\paragraph{Opacity regularization loss.}
As explained in Section~\ref{subsec:training}, the pretrained image diffusion model tends to generate simple and common shapes due to its learned biases.
In Figure~\ref{fig:ablation_reg_loss} (f), we show that when armrests and stretchers are not explicitly mentioned in the text prompt, images sampled from MVDream often lack these structures, leading to missing parts in the generated 3D shapes, even though the parts exist in the input coarse voxel grid; see Figure~\ref{fig:ablation_reg_loss} (a) and (b).
Since our model is trained on shapes with diverse structures, such parts should be handled automatically by the network without explicit text descriptions. Therefore, we use the regularization loss defined on the rendered masks to ensure structural consistency.

Figure~\ref{fig:ablation_reg_loss} (b-e) show qualitative results using the regularization loss with different $\lambda_{reg}$ values. By using a larger $\lambda_{reg}$, the generated shape can faithfully preserve the structure of the coarse voxels, even when the structure is asymmetric, e.g., with different numbers of stretchers and varying armrest lengths on each side. This is also reflected by the higher IoU in Table~\ref{tab:reg_ablation}. Note that the Render-FID of the model without the regularization loss, i.e. $\lambda_{reg}=0$, is slightly higher, since the generated shapes are more aligned with the preferences of the image diffusion model.
\paragraph{Two-stage learning.}
As discussed in Section ~\ref{sec:method_curriculum}, the image diffusion model's learned prior exhibits strong biases toward common structures. In Figure~\ref{fig:ablation_initialization} (b), when the text prompt indicates that the shape is to be a cake, despite that the input coarse shape is not round, the model attempts to generate a round cake, although a quarter of the cake is cut off by our masking (see Section ~\ref{sec:method_net}). By using our two-stage training strategy, the model effectively preserves the structure of the input coarse voxel grids while maintaining high-quality geometry and texture.

\subsection{Applications}
\label{sec:application}

\paragraph{Cross-category detailization.}
By using a highly generic text prompt that can describe multiple shape categories, such as ``furniture'' for chairs, tables, couches, and beds, our detailizer can generate a diverse collection of detailed 3D shapes across these categories, while maintaining consistent styles throughout, as shown in Figure ~\ref{fig:teaser}, where we have trained a single model using the coarse voxel grids from all four categories. Note that we did not provide any class labels to the network during training. Yet once trained, our detailizer can automatically identify the category of the input coarse voxel grid and upsample it into the corresponding detailed 3D shape.

\paragraph{Interactive modeling.}
Leveraging the feed-forward model design, our detailizer is capable of upsampling an input coarse voxel grid into a detailed 3D shape in under one second, making it possible to be incorporated into an interactive modeling application where users can refine and iterate on designs efficiently; see Figure~\ref{fig:procedural_editing}. We have developed an interactive modeling interface that enables users to edit a coarse voxel grid, select a text prompt, and visualize the resulting detailed and textured 3D shape; see supplementary video.

\section{Conclusions}
\label{sec:conclusions}





\begin{figure}
\begin{picture}(244, 94)
  \put(0, 0){\includegraphics[width=1\linewidth]{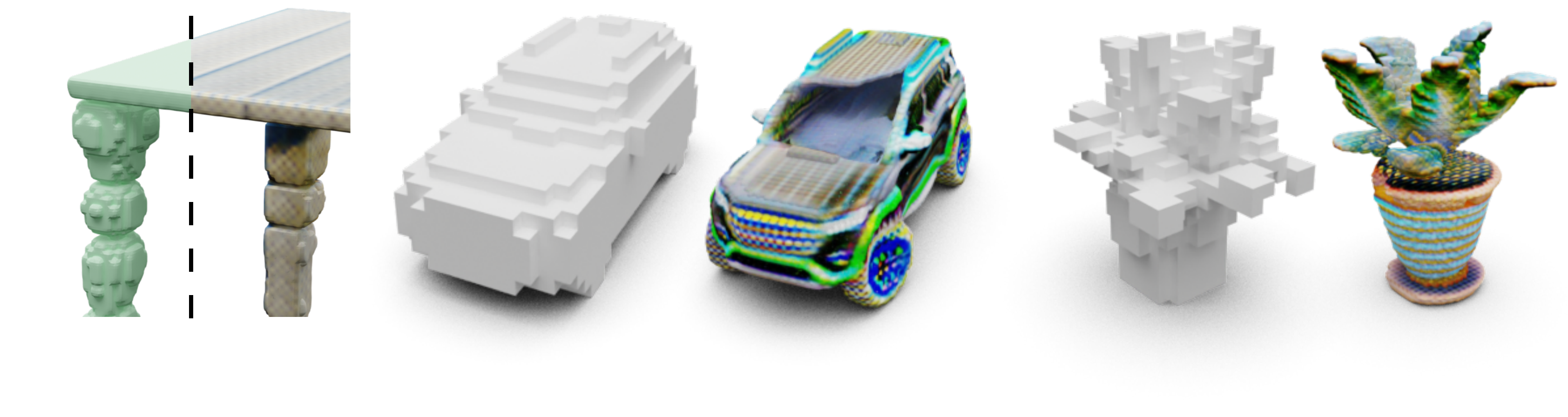}}
  \put(15, 3){\small (a) Table}
  \put(95, 3){\small (b) Car}
  \put(191, 3){\small (c) Plant}

  \put(5, 84){\small \textit{``a stone table with a}}
  \put(0, 74){\small \textit{heavy base and texture''}}

  \put(80, 84){\small \textit{``a modern SUV with a sturdy}}
  \put(85, 74){\small \textit{frame and tinted windows''}}

  \put(180, 84){\small \textit{``an asplenium nidus}}
  \put(180, 74){\small \textit{planted in a clay pot''}}

\end{picture}
\vspace{-7mm}
  \caption{Limitations. Our current representation is constrained in voxel resolution and is not suitable for generating shapes with glossy/reflective material or thin surfaces/threads. (a) left shows the geometry of the shape.}
  \vspace{-4mm}
  \label{fig:limitation}
\end{figure}

We introduce ART-DECO, a shape detailization model that can refine a coarse shape of arbitrary structure into a detailed 3D shape.
By distilling the knowledge in a pretrained multi-view image diffusion model, our detailizer learns to generate local details with the style described in an input text prompt.
Once trained, our detailizer can be used to detailize given coarse shapes into detailed shapes with a consistent style.
We demonstrate the superior performance of our method in the experiments and in an interactive modeling workflow.

\paragraph{Limitation.}
The dense voxel representation that our method adopts enables us to take advantage of convolutional networks. However, it limits the resolution of our generated shapes; the surfaces can appear blocky or jagged due to voxel artifacts, causing local details and smooth curves to be less faithfully presented. It also struggles with thin structures such as plant leaves; see Figure~\ref{fig:limitation} (a) and (c). Alternative representations and network design can be explored to address this issue. Our representation stores a fixed color in each voxel instead of view-dependent colors, therefore it cannot represent shapes with glossy or reflective surfaces, see Figure~\ref{fig:limitation} (b). Our detailizer requires re-training for each different text prompt, which can be inconvenient for fast prototyping. Lastly, our method requires a set of category-specific coarse shapes for training each new style, which limits its applicability in domains where such data does not exist, and makes it less feasible to scale across many styles without additional human effort. With sufficient compute, we believe it is possible to distill a more generic detailizer that can take both the text prompt and the coarse shape as input during inference, and generate detailed shapes at an interactive speed.
\section*{Acknowledgments}

This work was done during the first author's internship at Adobe Research, and it is supported in part by an NSERC grant (No. 611370) and a gift fund from Adobe Research.

\bibliographystyle{ACM-Reference-Format}
\bibliography{bibliography}

\clearpage
\begin{figure*}
\vspace{-0mm}
\begin{picture}(510, 54)
\centering
  \put(0, 0){\includegraphics[width=1.0\linewidth]{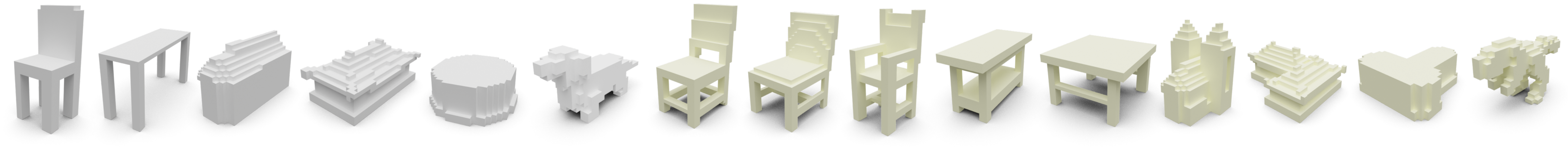}}
  
\end{picture}
\vspace{-9mm}
  \caption{Training set examples of coarse shapes: simple structures are shown in white, and complex ones in light yellow.}
  \label{fig:simple_complex_coarse}
\end{figure*}

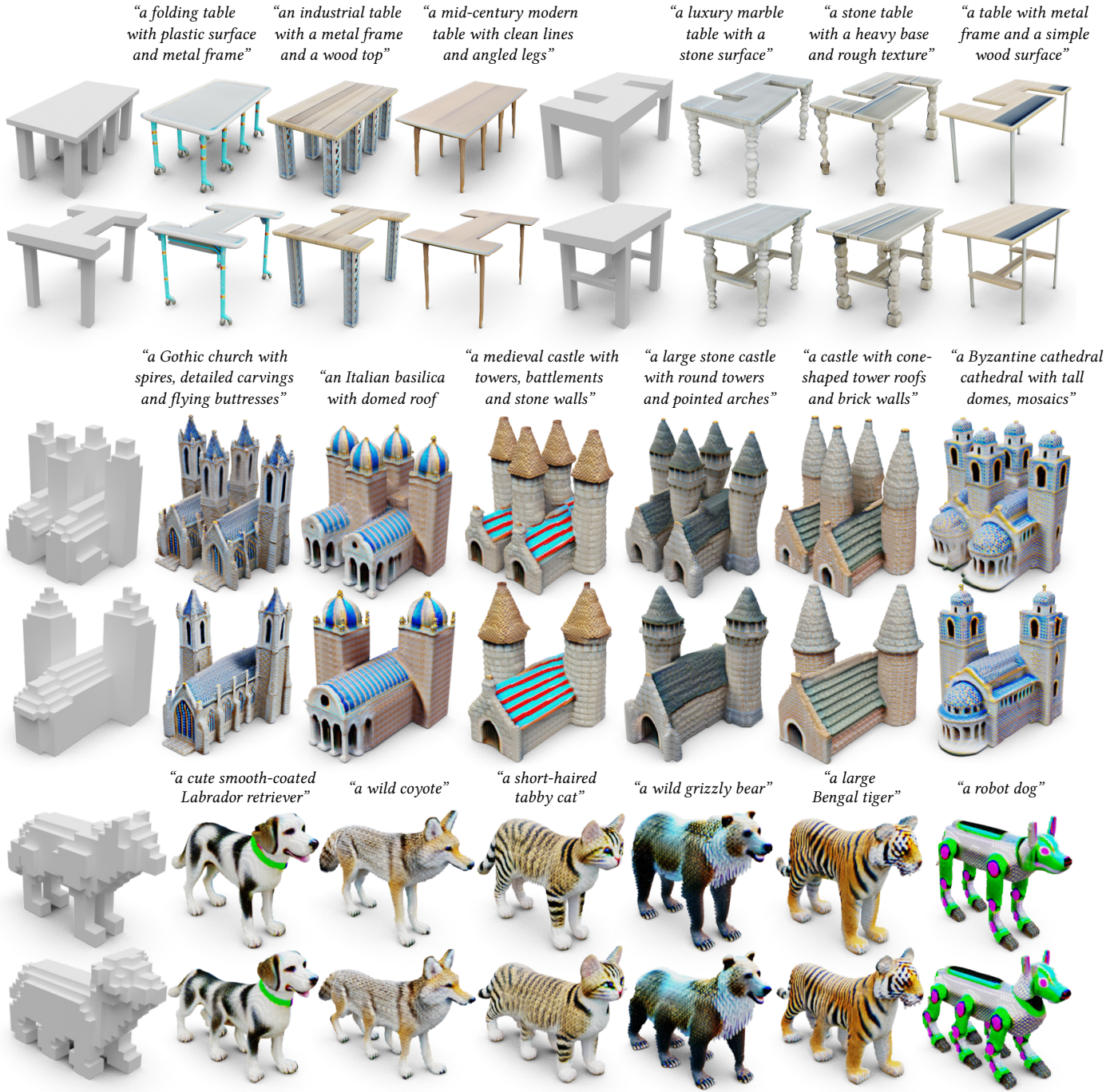
\begin{figure*}
\begin{picture}(510, 527)
\centering
  \put(0, 0){\includegraphics[width=1.0\linewidth]{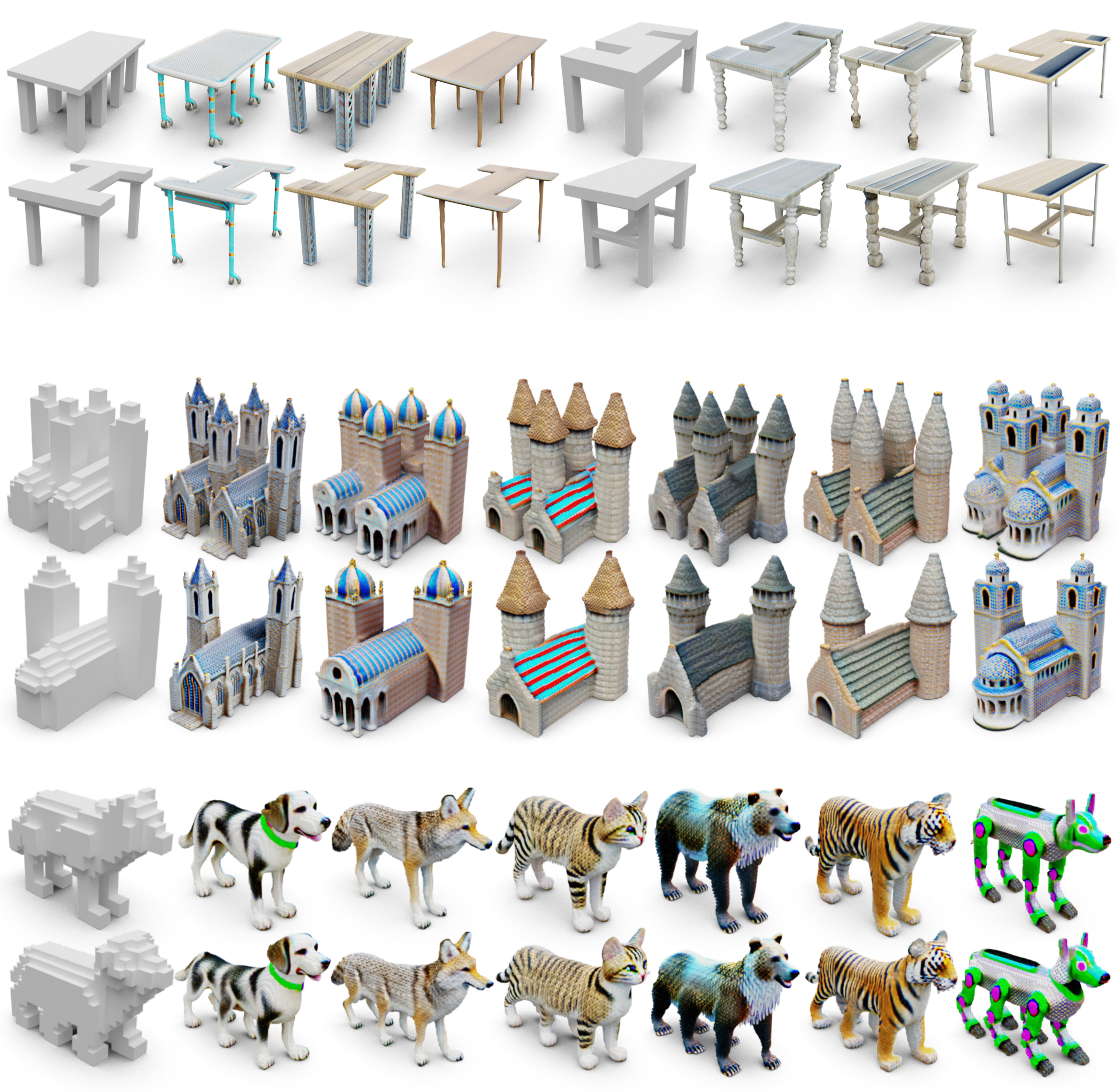}}

  \put(80, 152){\small \textit{``a cute smooth-coated}}
  \put(86, 143){\small \textit{Labrador retriever''}}

  \put(165, 147){\small \textit{``a wild coyote''}}

  \put(235, 152){\small \textit{``a short-haired}}
  \put(244, 143){\small \textit{tabby cat''}}

  \put(300, 147){\small \textit{``a wild grizzly bear''}}

  \put(390, 152){\small \textit{``a large}}
  \put(385, 143){\small \textit{Bengal tiger''}}

  \put(453, 147){\small \textit{``a robot dog''}}
  \put(67, 352){\small \textit{``a Gothic church with}}
  \put(64, 342){\small \textit{spires, detailed carvings}}
  \put(67, 332){\small \textit{and flying buttresses''}}

  \put(151, 347){\small \textit{``an Italian basilica}}
  \put(155, 337){\small \textit{with domed roof}}

  \put(220, 352){\small \textit{``a medieval castle with}}
  \put(225, 342){\small \textit{towers, battlements}}
  \put(230, 332){\small \textit{and stone walls''}}

  \put(305, 352){\small \textit{``a large stone castle}}
  \put(305, 342){\small \textit{with round towers}}
  \put(305, 332){\small \textit{and pointed arches''}}

  \put(380, 352){\small \textit{``a castle with cone-}}
  \put(380, 342){\small \textit{shaped tower roofs}}
  \put(383, 332){\small \textit{and brick walls''}}

  \put(450, 352){\small \textit{``a Byzantine cathedral}}
  \put(455, 342){\small \textit{cathedral with tall}}
  \put(458, 332){\small \textit{domes, mosaics''}}
  \put(63, 515){\small \textit{``a folding table}}
  \put(60, 505){\small \textit{with plastic surface}}
  \put(63, 495){\small \textit{and metal frame''}}

  \put(129, 515){\small \textit{``an industrial table}}
  \put(130, 505){\small \textit{with a metal frame}}
  \put(133, 495){\small \textit{and a wood top''}}

  \put(200, 515){\small \textit{``a mid-century modern}}
  \put(205, 505){\small \textit{table with clean lines}}
  \put(210, 495){\small \textit{and angled legs''}}

  \put(317, 515){\small \textit{``a luxury marble}}
  \put(325, 505){\small \textit{table with a}}
  \put(322, 495){\small \textit{stone surface''}}

  \put(390, 515){\small \textit{``a stone table}}
  \put(383, 505){\small \textit{with a heavy base}}
  \put(383, 495){\small \textit{and rough texture''}}

  \put(455, 515){\small \textit{``a table with metal}}
  \put(455, 505){\small \textit{frame and a simple}}
  \put(462, 495){\small \textit{wood surface''}}
  
\end{picture}
\vspace{-9mm}
  \caption{Results of text-guided detailization with input coarse voxels control. We show the input coarse voxels on the left and the text prompts on the top. Geometry-only visualization can be found in the supplementary.}
  \label{fig:additional_results_2_1}
\end{figure*}
\clearpage
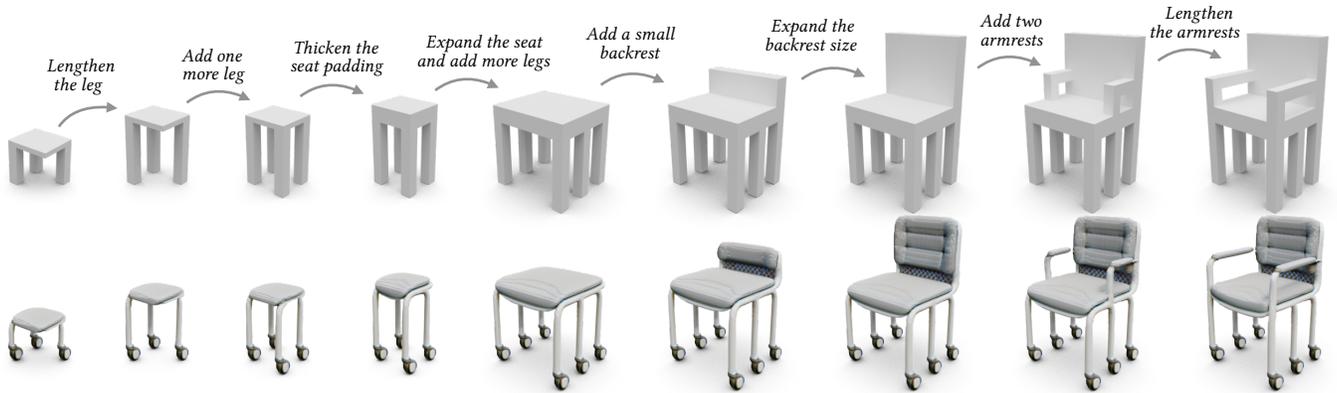
\begin{figure*}
\begin{picture}(510, 160)
\centering
  \put(0, 0){\includegraphics[width=1.0\linewidth]{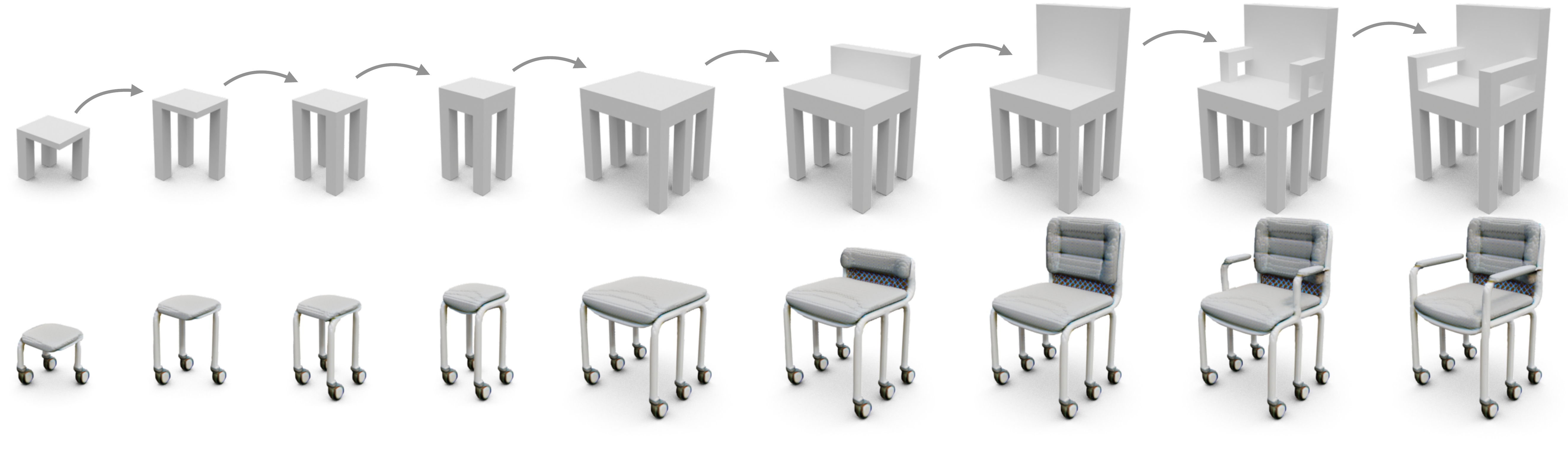}}

  \put(20, 135){\footnotesize \textit{Lengthen}}
  \put(22.5, 128){\footnotesize \textit{the leg}}

  \put(71, 139){\footnotesize \textit{Add one}}
  \put(71, 132){\footnotesize \textit{more leg}}

  \put(113, 142){\footnotesize \textit{Thicken the}}
  \put(112, 135){\footnotesize \textit{seat padding}}

  \put(163.5, 144){\footnotesize \textit{Expand the seat}}
  \put(160, 137){\footnotesize \textit{and add more legs}}

  \put(224, 147){\footnotesize \textit{Add a small}}
  \put(229, 140){\footnotesize \textit{backrest}}

  \put(294, 150){\footnotesize \textit{Expand the}}
  \put(291.5, 143){\footnotesize \textit{backrest size}}

  \put(373, 152){\footnotesize \textit{Add two}}
  \put(373, 145){\footnotesize \textit{armrests}}

  \put(443, 155){\footnotesize \textit{Lengthen}}
  \put(437, 148){\footnotesize \textit{the armrests}}
  
\end{picture}
\vspace{-11mm}
  \caption{Example of procedural editing. After training the model with the text prompt \textit{``an office chair with wheels and thick padding''}, the detailization of each edit takes \textbf{less than one second}. Our method demonstrates strong robustness to minor modifications and local edits.}
  \label{fig:procedural_editing}
\end{figure*}

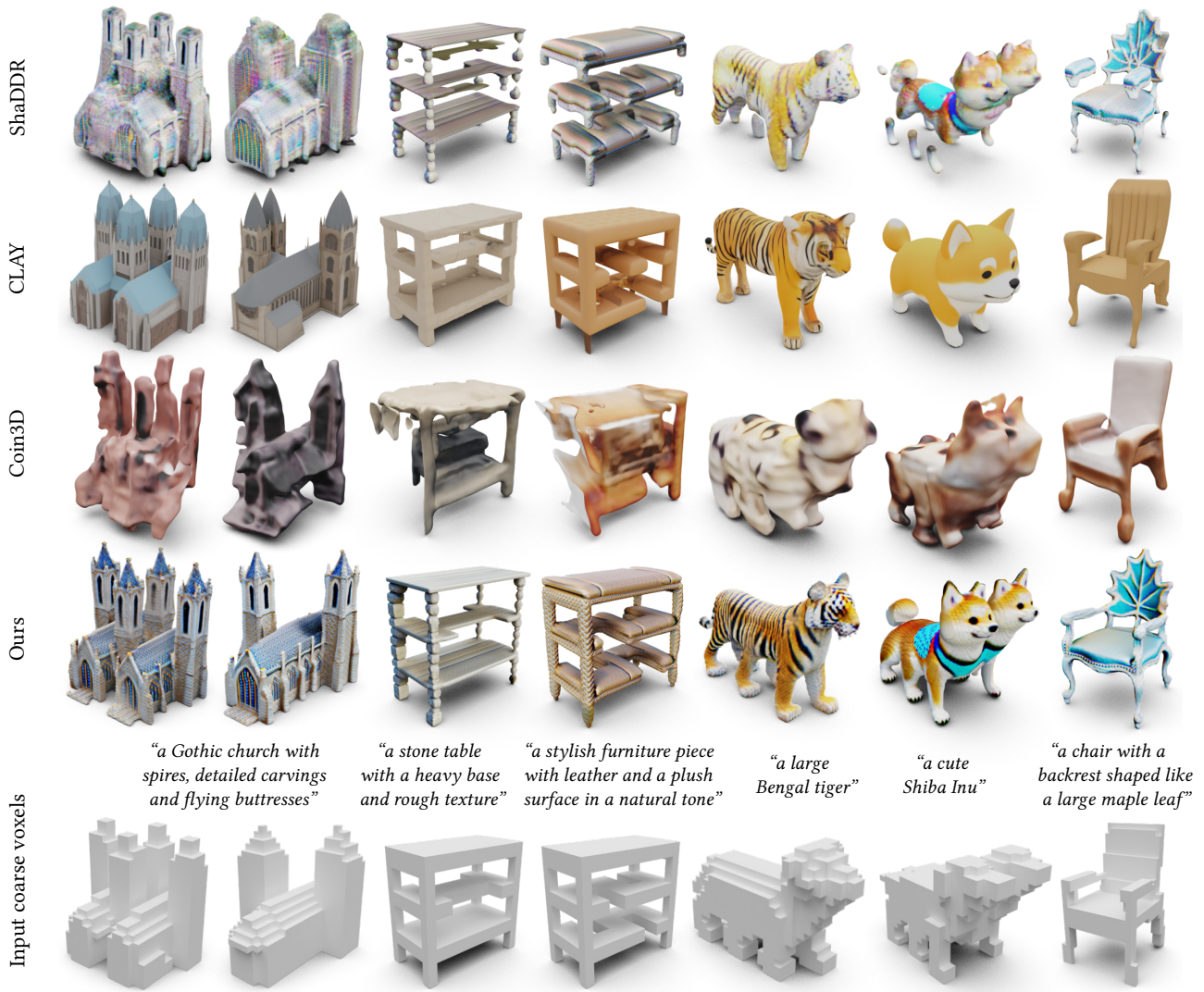
\begin{figure*}
\begin{picture}(510, 407)
\centering
  \put(30, 0){\includegraphics[width=0.9\linewidth]{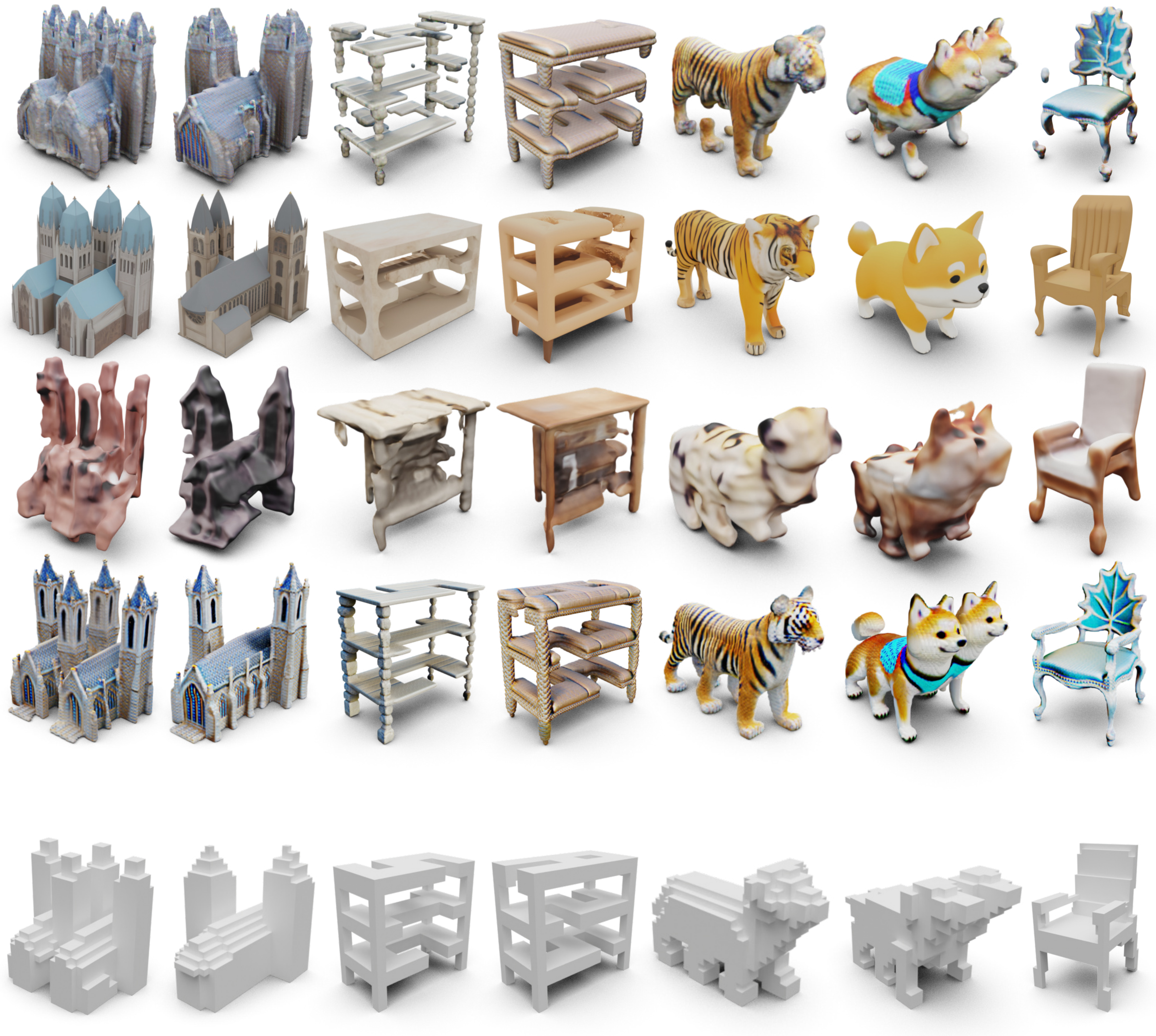}}

  \put(10, 20){\text{\rotatebox{90}{Input coarse voxels}}}
  \put(10, 145){\text{\rotatebox{90}{Ours}}}
  \put(10, 220){\text{\rotatebox{90}{Coin3D}}}
  \put(10, 295){\text{\rotatebox{90}{CLAY}}}
  \put(10, 360){\text{\rotatebox{90}{ShaDDR}}}

  \put(67, 106){\small \textit{``a Gothic church with}}
  \put(64, 96){\small \textit{spires, detailed carvings}}
  \put(67, 86){\small \textit{and flying buttresses''}}

  \put(160, 106){\small \textit{``a stone table}}
  \put(153, 96){\small \textit{with a heavy base}}
  \put(153, 86){\small \textit{and rough texture''}}

  \put(220, 106){\small \textit{``a stylish furniture piece}}
  \put(220, 96){\small \textit{with leather and a plush}}
  \put(220, 86){\small \textit{surface in a natural tone''}}

  \put(320, 101){\small \textit{``a large}}
  \put(315, 91){\small \textit{Bengal tiger''}}

  \put(380, 101){\small \textit{``a cute}}
  \put(375, 91){\small \textit{Shiba Inu''}}

  \put(435, 106){\small \textit{``a chair with a}}
  \put(430, 96){\small \textit{backrest shaped like}}
  \put(432, 86){\small \textit{a large maple leaf''}}
\end{picture}
\vspace{-9mm}
  \caption{Qualitative comparison of text-guided detailization with input coarse voxels control. Each layer of the two coarse tables is structured to represent letters \textit{\textbf{``S'', ``I'', ``G'', ``A'', ``2'', ``5''}}. Note that even with a simple structure, as shown in the last column, existing methods fail to produce plausible results when given a creative text prompt.}
  \label{fig:additional_results_1_1}
\end{figure*}

\twocolumn[
\centering
\textbf{ART-DECO: \underline{Ar}bitrary \underline{T}ext Guidance for 3D \underline{De}tailizer \underline{Co}nstruction} \\
\vspace{0.5em}\Large{(Supplementary Material)} \\
\vspace{1.0em}
]
\setcounter{page}{1}

\appendix

\section{Data, text prompts and code}

We evaluate our model in seven shape categories: chair, table, couch, bed, building, animal, and cake. We collect 1,505 chairs, 1,292 tables, 300 couches and 300 beds from ShapeNet~\cite{chang2015shapenet}, 32 buildings and 16 cakes from 3D Warehouse~\cite{3dwarehouse} under CC-BY 4.0, and 2,082 animals from Objaverse~\cite{deitke2023objaverse}. We voxelize the 3D shapes as our training set. For building and cake, we perform data augmentation to obtain 1,200 voxel grids for each category as our training set.

We use the following text prompts for text-guided detailization.
\begin{itemize}
    \item \textit{``a farmhouse chair with a cross-back design and a natural wood finish''}
    \item \textit{``a Queen Anne chair with a leather back in a light, neutral tone and a cushioned seat''}
    \item \textit{``a metal chair with a leather back and a cushioned seat''}
    \item \textit{``a chair with a backrest shaped like a large maple leaf''}
    \item \textit{``a Scandinavian-style chair with a clean design and soft fabric padding''}
    \item \textit{``an office-style chair with wheels and a thick seat padding''}
    \item \textit{``a rustic wooden chair with a rough texture and a simple, handcrafted look''}
    \item \textit{``a Victorian chair with elegant curves and velvet upholstery''}
    \item \textit{``a traditional Japanese palace with tiled roofs and wooden walls''}
    \item \textit{``a Gothic church with spires, detailed carvings, and flying buttresses''}
    \item \textit{``a Russian cathedral with domes and tall spires''}
    \item \textit{``a large stone castle with round towers and pointed arches''}
    \item \textit{``an Italian basilica with domed roof and arched windows''}
    \item \textit{``a castle with cone-shaped tower roofs and brick walls''}
    \item \textit{``a stone cathedral with a tall rose window and flying arches''}
    \item \textit{``an old German cathedral with timber framing and steep roof''}
    \item \textit{``a Gothic church with a bell tower and steep tiled roof''}
    \item \textit{``a medieval castle with towers, battlements, and stone walls''}
    \item \textit{``a Byzantine cathedral with tall domes, mosaics''}
    \item \textit{``a table with a metal frame and a simple wood surface''}
    \item \textit{``a luxury marble table with a stone surface''}
    \item \textit{``a stone table with a heavy base and rough texture''}
    \item \textit{``a folding table with a plastic surface and metal frame''}
    \item \textit{``a classic Queen Anne table with a smooth and muted finish''}
    \item \textit{``a mid-century modern table with clean lines and angled legs''}
    \item \textit{``an industrial style table with a metal frame and a thick wood top''}
    \item \textit{``a cute golden retriever''}
    \item \textit{``a cute bulldog''}
    \item \textit{``a cute smooth-coated Labrador retriever''}
    \item \textit{``a cute Shiba Inu''}
    \item \textit{``a wild coyote''}
    \item \textit{``a wild grizzly bear''}
    \item \textit{``a large Bengal tiger''}
    \item \textit{``a robot dog''}
    \item \textit{``a cake with chocolate dripping down the sides''}
\end{itemize}

We use the following prompts for cross-category detailization.
\begin{itemize}
    \item \textit{``a classic furniture piece made of polished wood with subtle details''}
    \item \textit{``an old Queen Anne style furniture in a light, neutral tone''}
    \item \textit{``a stylish furniture piece featuring leather and a plush surface in a natural tone''}
\end{itemize}

We will provide the ready-to-use data and code upon publication.

\section{Network architecture}

We took inspiration from ShaDDR~\cite{chen2024shadder}, where the input voxel grid is dilated by one voxel and two branches of networks are used to generate geometry and texture, respectively. For the density upsampling network, we use 5 layers of 3D convolution to extract the features of the input coarse voxel grid, followed by 2 layers of transposed 3D convolution for upsampling. Each upsampling layer doubles the input resolution. For the albedo upsampling network, we use the same network architecture as the density upsampling network, except that the final upsampling layer outputs three channels to represent RGB values.

We use a learning rate of $10^{-4}$, a batch size of $1$, and the Adam optimizer for all experiments.

\section{Evaluation metrics}

We use Render-FID and CLIP score to quantitatively evaluate the quality of the generated shapes from the rendering perspective. We also use Strict-IoU and Loose-IoU to quantitatively evaluate the structure consistency of the generated shapes. For the chair category, we use the test set from DECOLLAGE. For building and cake categories, we randomly augment voxels to create the test set. 

Generating a single shape using CLAY's commercial product, Rodin, takes approximately 5 to 10 minutes, while Coin3D requires about 25 minutes to optimize one shape, which makes both approaches impractical for large-scale automatic evaluation.
Therefore, we randomly select 8 text prompts. For each text prompt, we randomly sample 10 coarse
voxels from the corresponding test set to compute the metrics.

\paragraph{Render-FID} For each generated detailed shape, we uniformly render 4 views at 0, 90, 180, and 270 degrees. For each text prompt, we first augment it with view-dependent description by appending \textit{front view, side view, back view} to the end of the prompt. For each augmented text prompt, We then use Stable Diffusion 2.1 with \textit{stabilityai/stable-diffusion-2-1-base} to generate 4 different images. We compute the FID between the rendered images of the generated shapes and the images generated from Stable Diffusion.

\paragraph{CLIP score} For each generated detailed shape, we uniformly render 24 views by rotating the camera around the object with fixed poses. We then compute the cosine similarity between the CLIP embeddings of the input text prompt and the rendered images of the generated shape. We average the score over all the views. We compute the CLIP score using \textit{openai/clip-vit-large-patch14} version of CLIP model.
\begin{align}
    \text{CLIPScore} = max(100\cdot cos(E_{1}, E_{2}), 0)
\end{align}

\paragraph{Strict-IoU} Given the density field of the generated shape $v_d$ and the corresponding input coarse voxel grid $v$, we first voxelize the density field using a threshold value of $30$. We then downsample it to the same resolution as the input coarse voxel grid, which we define as $v'$. We compute the Strict-IoU as follows:
\begin{align}
    \text{Strict-IoU} = \frac{||v \; \& \; v'||_{1}}{||v \; | \; v'||_{1}}
\end{align}

\paragraph{Loose-IoU} Loose-IoU is the relaxed version of Strict-IoU and we define it as follows:
\begin{align}
    \text{Loose-IoU} = \frac{||v \; \& \; v'||_{1}}{||v||_{1}}
\end{align}

\section{Coin3D implementation}

We use the teddy bear example included in the officially released code to demonstrate that Coin3D can work on simple structures. Figure~\ref{fig:coin3d_intermediate_final_result_repro} shows the intermediate results and the final output generated by the officially released code. We also show the intermediate results and the final output of Coin3D when applied to the complex structure in Figure~\ref{fig:coin3d_intermediate_complex_structure}. Coin3D performs well on simple structures but struggles to handle more complex ones.

\section{ShaDDR implementation}

We use the original implementation of ShaDDR and its default settings: an input resolution of $k=32$ and an output resolution of $K=256$. We provide the detailed shapes generated by our method as styles for training ShaDDR in Figure~\ref{fig:shaddr_styles}.

\section{Geometry-only visualizations}

\begin{figure*}
\begin{picture}(510, 565)
\centering
  \put(0, 0){\includegraphics[width=1.0\linewidth]{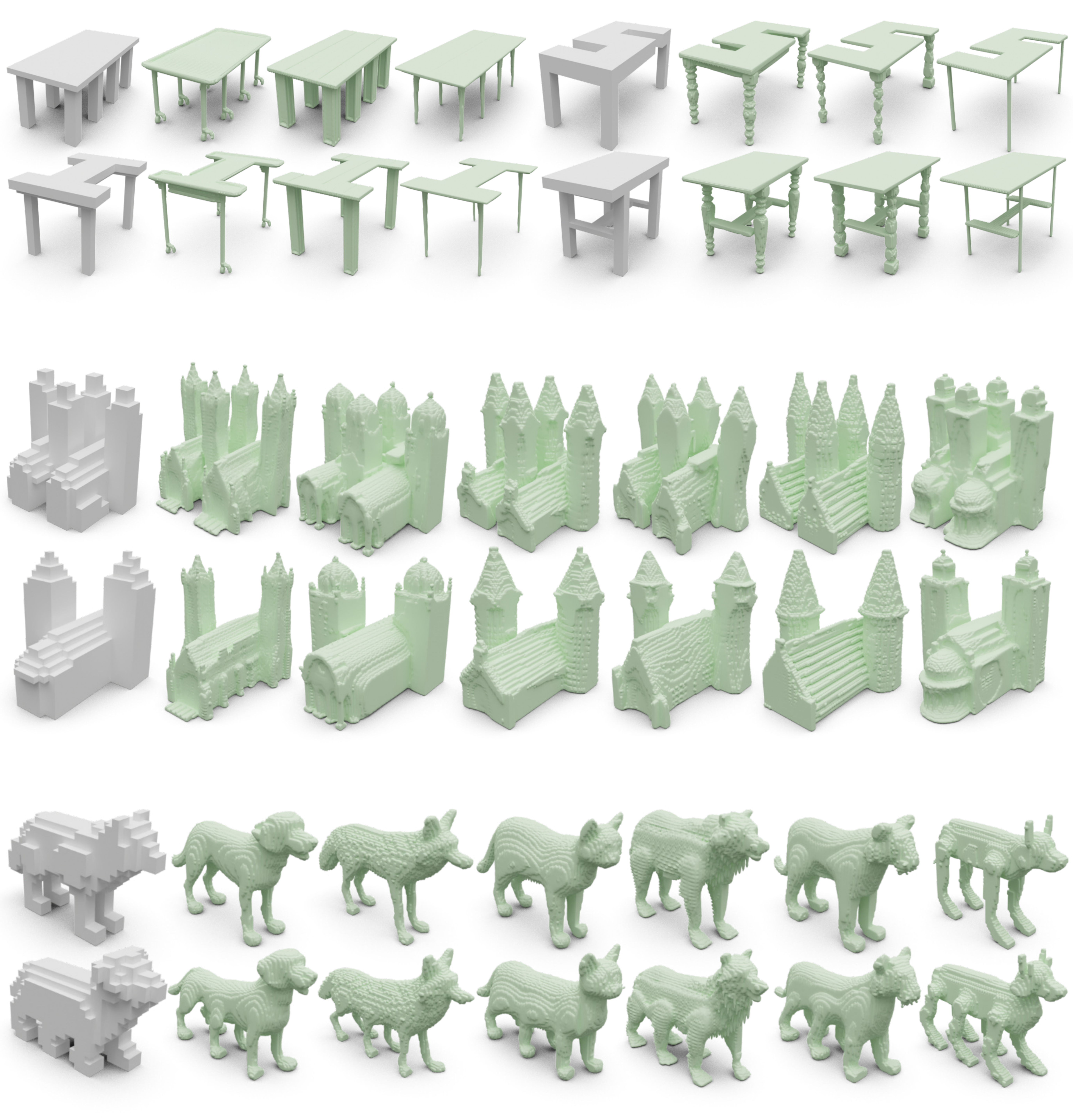}}

  \put(80, 167){\small \textit{``a cute smooth-coated}}
  \put(86, 157){\small \textit{Labrador retriever''}}

  \put(165, 162){\small \textit{``a wild coyote''}}

  \put(235, 167){\small \textit{``a short-haired}}
  \put(244, 157){\small \textit{tabby cat''}}

  \put(300, 162){\small \textit{``a wild grizzly bear''}}

  \put(390, 167){\small \textit{``a large}}
  \put(385, 157){\small \textit{Bengal tiger''}}

  \put(453, 162){\small \textit{``a robot dog''}}
  \put(67, 387){\small \textit{``a Gothic church with}}
  \put(64, 377){\small \textit{spires, detailed carvings}}
  \put(67, 367){\small \textit{and flying buttresses''}}

  \put(151, 382){\small \textit{``an Italian basilica}}
  \put(155, 372){\small \textit{with domed roof}}

  \put(220, 387){\small \textit{``a medieval castle with}}
  \put(225, 377){\small \textit{towers, battlements}}
  \put(230, 367){\small \textit{and stone walls''}}

  \put(305, 387){\small \textit{``a large stone castle}}
  \put(305, 377){\small \textit{with round towers}}
  \put(305, 367){\small \textit{and pointed arches''}}

  \put(380, 387){\small \textit{``a castle with cone-}}
  \put(380, 377){\small \textit{shaped tower roofs}}
  \put(383, 367){\small \textit{and brick walls''}}

  \put(450, 387){\small \textit{``a Byzantine cathedral}}
  \put(455, 377){\small \textit{cathedral with tall}}
  \put(458, 367){\small \textit{domes, mosaics''}}
  \put(63, 550){\small \textit{``a folding table}}
  \put(60, 540){\small \textit{with plastic surface}}
  \put(63, 530){\small \textit{and metal frame''}}

  \put(129, 550){\small \textit{``an industrial table}}
  \put(130, 540){\small \textit{with a metal frame}}
  \put(133, 530){\small \textit{and a wood top''}}

  \put(200, 550){\small \textit{``a mid-century modern}}
  \put(205, 540){\small \textit{table with clean lines}}
  \put(210, 530){\small \textit{and angled legs''}}

  \put(317, 550){\small \textit{``a luxury marble}}
  \put(325, 540){\small \textit{table with a}}
  \put(322, 530){\small \textit{stone surface''}}

  \put(390, 550){\small \textit{``a stone table}}
  \put(383, 540){\small \textit{with a heavy base}}
  \put(383, 530){\small \textit{and rough texture''}}

  \put(455, 550){\small \textit{``a table with metal}}
  \put(455, 540){\small \textit{frame and a simple}}
  \put(462, 530){\small \textit{wood surface''}}
  
\end{picture}
\vspace{-9mm}
  \caption{Geometry-only visualization of text-guided detailization with input coarse voxels control. We show the input coarse voxels on the left and the text prompts on the top.}
  \label{fig:additional_results_2_1_geo_only}
\end{figure*}

Figure~\ref{fig:additional_results_2_1_geo_only} shows the geometry-only visualizations corresponding to Figure 12 in the main paper.

\section{Additional qualitative comparisons}

\begin{figure*}
\begin{picture}(510, 620)
\centering
  \put(50, 0){\includegraphics[width=0.8\linewidth]{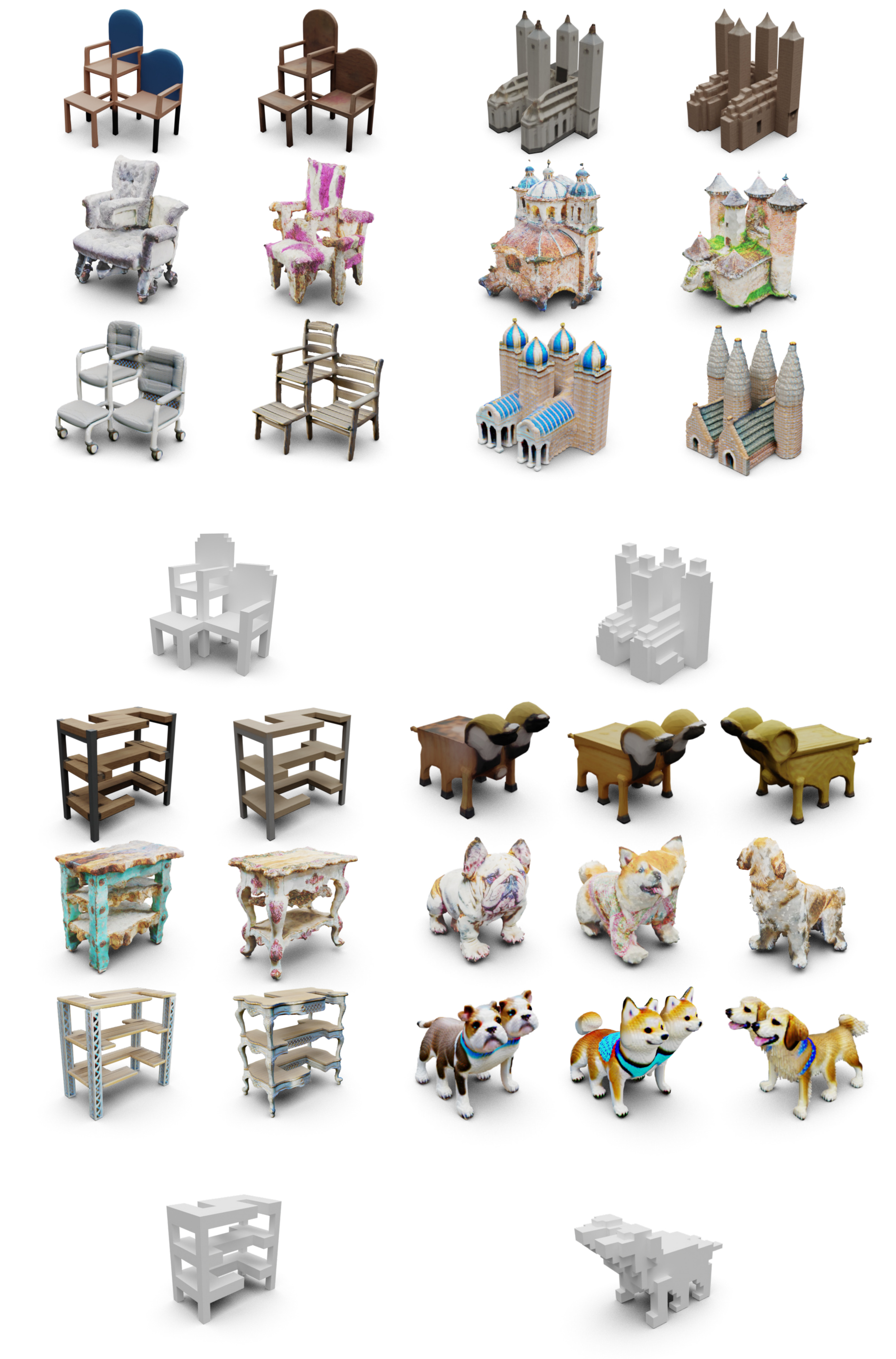}}

  \put(30, 130){\text{\rotatebox{90}{Ours}}}
  \put(30, 185){\text{\rotatebox{90}{Sherpa3D}}}
  \put(30, 255){\text{\rotatebox{90}{Trellis}}}

  \put(30, 430){\text{\rotatebox{90}{Ours}}}
  \put(30, 495){\text{\rotatebox{90}{Sherpa3D}}}
  \put(30, 570){\text{\rotatebox{90}{Trellis}}}

  \put(50, 95){\small \textit{``an industrial table with a}}
  \put(50, 85){\small \textit{metal frame and wood top''}}

  \put(150, 95){\small \textit{``a Queen Anne table with}}
  \put(150, 85){\small \textit{smooth and muted finish''}}

  \put(250, 90){\small \textit{``a cute bulldog''}}
  \put(310, 90){\small \textit{``a cute Shiba Inu''}}
  \put(380, 90){\small \textit{``a cute golden retriever''}}

  \put(73, 395){\small \textit{``an office chair with}}
  \put(65, 385){\small \textit{wheels and thick padding''}}

  \put(160, 395){\small \textit{``a rustic wooden chair}}
  \put(160, 385){\small \textit{with rough texture look''}}

  \put(265, 395){\small \textit{``an Italian basilica}}
  \put(267, 385){\small \textit{with domed roof''}}

  \put(350, 395){\small \textit{``a castle with cone-shaped}}
  \put(350, 385){\small \textit{tower roofs and brick walls''}}

\end{picture}
\vspace{-10mm}
  \caption{Additional comparison results with TRELLIS and Sherpa3D.}
  \label{fig:trellis_sherpa3d_fig}
\end{figure*}

\begin{figure*}
\begin{picture}(510, 580)
\centering
  \put(50, 0){\includegraphics[width=0.8\linewidth]{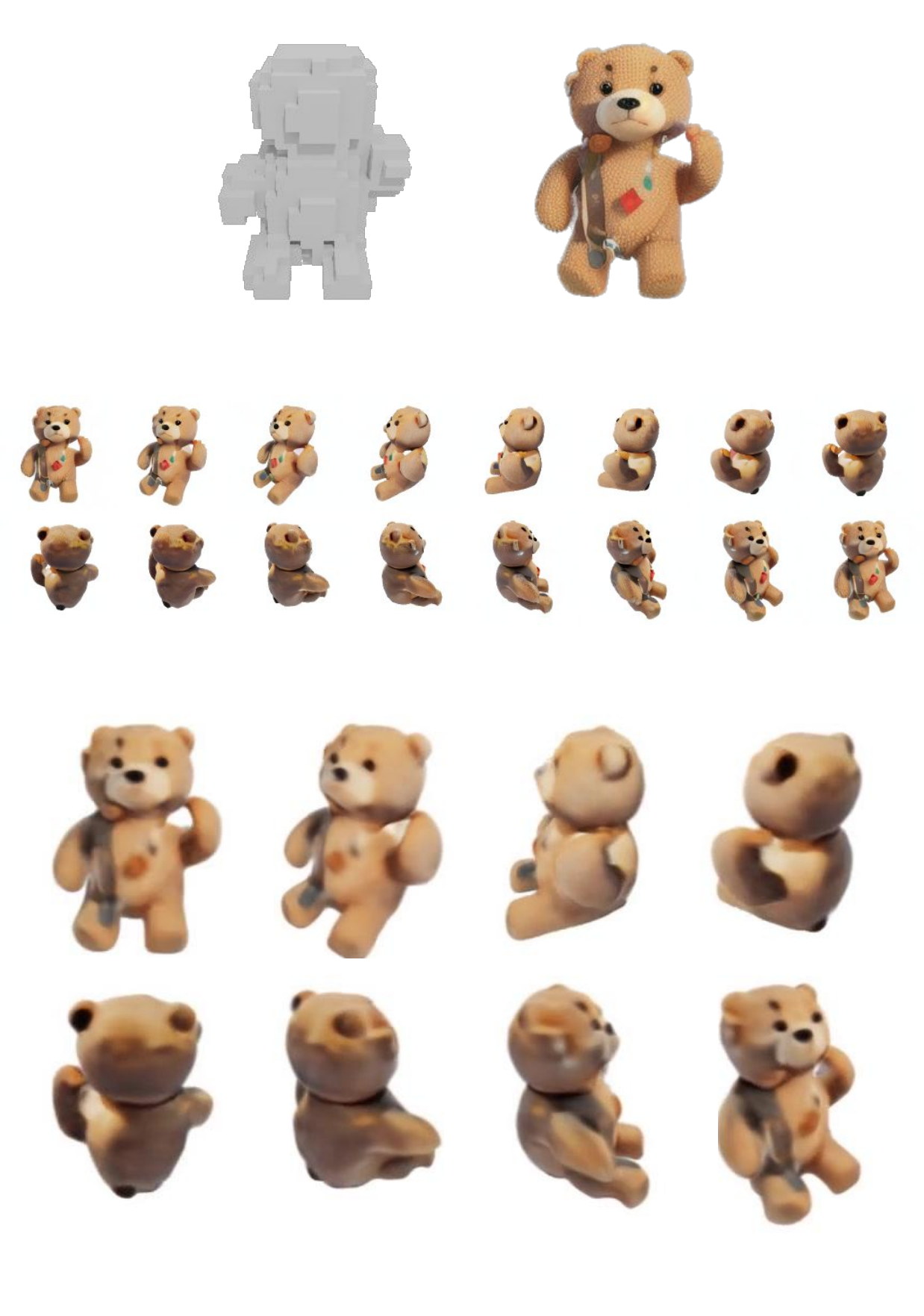}}

  \put(95, 420){(a) Rendering of the input coarse proxy}
  \put(290, 420){(b) Single-view image}
  
  \put(220, 285){(c) Multi-view images}

  \put(180, 20){(d) Rendering of NeuS reconstruction}
\end{picture}
\vspace{-7mm}
  \caption{Coin3D reproduction experiment with teddy bear coarse proxy and text prompt \textit{"a lovely teddy bear"}. We show (a) the rendering of the input coarse proxy, (b) the single-view image generated by 2D ControlNet conditioned on the text prompt \textit{a lovely teddy bear}, (c) the generated multi-view images and (d) the rendered images of NeuS reconstruction.}
  \label{fig:coin3d_intermediate_final_result_repro}
\end{figure*}

\begin{figure*}
\begin{picture}(510, 580)
\centering
  \put(50, 0){\includegraphics[width=0.8\linewidth]{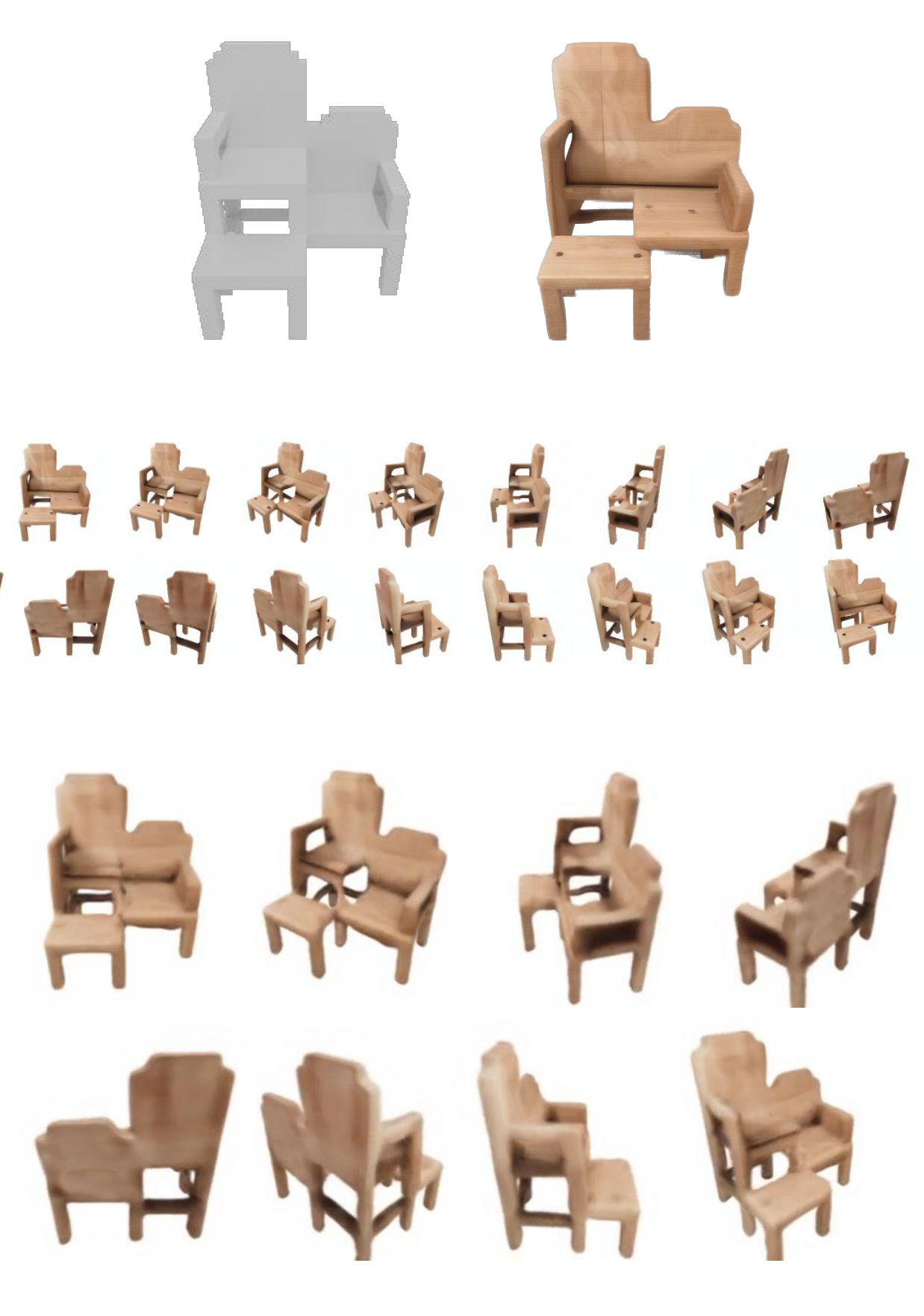}}

  \put(95, 410){(a) Rendering of the input coarse proxy}
  \put(290, 410){(b) Single-view image}
  
  \put(220, 260){(c) Multi-view images}

  \put(180, 0){(d) Rendering of NeuS reconstruction from (c)}
\end{picture}
\vspace{-3mm}
  \caption{Coin3D applied to a complex chair structure with text prompt \textit{"a chair with a backrest shaped like a large maple leaf"}. We show (a) the rendering of the input coarse proxy, (b) the single-view image generated by 2D ControlNet conditioned on the text prompt \textit{a lovely teddy bear}, (c) the generated multi-view images and (d) the rendered images of NeuS reconstruction. }
  \label{fig:coin3d_intermediate_complex_structure}
\end{figure*}

\begin{figure*}
\begin{picture}(510, 175)
\centering
  \put(0, 0){\includegraphics[width=1.0\linewidth]{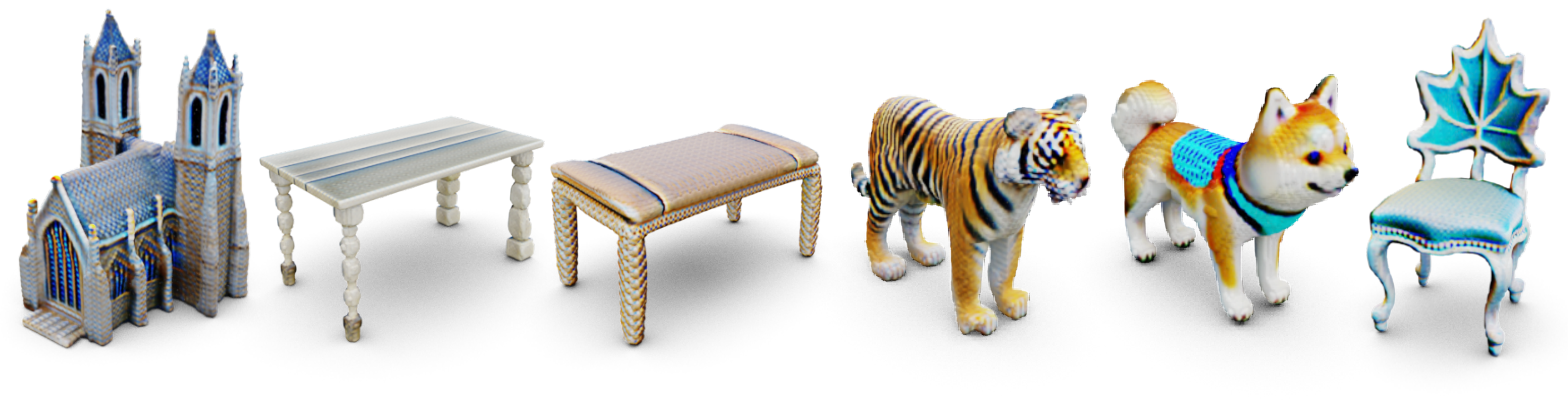}}

  \put(3, 160){\small \textit{``a Gothic church with}}
  \put(0, 150){\small \textit{spires, detailed carvings}}
  \put(3, 140){\small \textit{and flying buttresses''}}

  \put(110, 160){\small \textit{``a stone table}}
  \put(103, 150){\small \textit{with a heavy base}}
  \put(103, 140){\small \textit{and rough texture''}}

  \put(195, 160){\small \textit{``a stylish furniture piece}}
  \put(195, 150){\small \textit{with leather and a plush}}
  \put(195, 140){\small \textit{surface in a natural tone''}}

  \put(307, 150){\small \textit{``a large}}
  \put(302, 140){\small \textit{Bengal tiger''}}

  \put(388, 150){\small \textit{``a cute}}
  \put(383, 140){\small \textit{Shiba Inu''}}

  \put(455, 160){\small \textit{``a chair with a}}
  \put(450, 150){\small \textit{backrest shaped like}}
  \put(452, 140){\small \textit{a large maple leaf''}}

\end{picture}
\vspace{-10mm}
  \caption{We show the detailed shapes generated by our method, which are used as styles shapes for training ShaDDR. These detailed shapes are of good quality and suitable for training ShaDDR.}
  \label{fig:shaddr_styles}
\end{figure*}

Figure~\ref{fig:trellis_sherpa3d_fig} shows additional comparisons with Trellis and Sherpa3D. Trellis tends to preserve only the input coarse structure and fails to add geometric or texture details, resulting in outputs that do not match the style of the input text prompt. Sherpa3D tends to disregard the coarse initialization and instead optimizes the geometry to match the priors of its 2D diffusion model. For example, given a coarse structure of an animal with two heads and six legs, Sherpa3D simplifies it and generates one head and four legs. The structures of chairs, tables, and buildings are also overly simplified.

\section{User Study}

For the user study, we used the same examples that were used to compute the quantitative results. We designed 40 questions and collected responses from 42 participants. For each question, we show the participants an input coarse shape for structural control, a text prompt describing the style, and several 3D shapes generated by our method and three baselines. We ask the participants to select the BEST generated 3D shape based on several aspects:
\begin{itemize}
    \item Style adherence - How well the generated 3D shape follows the style described by the text.
    \item Quality of the generated details - As judged by the fine-grained details of the generated 3D shape.
    \item Structure preservation - How well the generated shape follows the structure of the input coarse shape.
    \item Overall preference - As judged by the overall impression of the generation, all factors considered.
\end{itemize}

We report the average percentage of participants selecting each method as producing the best results. As shown in Table~\ref{tab:user_study}, 83.45\% of the participants selected the shapes generated by our method as the most preferred result, highlighting its advantage over competing approaches.

\begin{table}[t]
    \begin{center}
    \caption{User study results. The average percentage of participants selecting each method as producing the best results.
    }
    \label{tab:user_study}
    \vspace{-3mm}
    \resizebox{1.0\columnwidth}{!}{
    \begin{tabular}{lcccc}
    \toprule
        & Ours & CLAY & Coin3D & ShaDDR \\
    \midrule
       Avg. preference & \textbf{83.45\%} & 15.77\% & 0.06\% & 0.72\% \\
    \bottomrule
    \end{tabular}
    }
    \end{center}
\end{table}

\section{Additional quantitative comparisons}

Table~\ref{tab:farmhouse_chair} to~\ref{tab:maple_leaf_chair} show the quantitative comparisons of randomly selected text prompts in our experiments.

\section{Additional quantitative ablations}

Table~\ref{tab:farmhouse_chair_ablation} to~\ref{tab:maple_leaf_chair_ablation} show the quantitative ablation of using regularization loss with different $\lambda_{reg}$ on randomly selected text prompts.

\section{Additional qualitative results}

\begin{figure*}
\begin{picture}(510, 400)
\centering
  \put(0, 0){\includegraphics[width=1.0\linewidth]{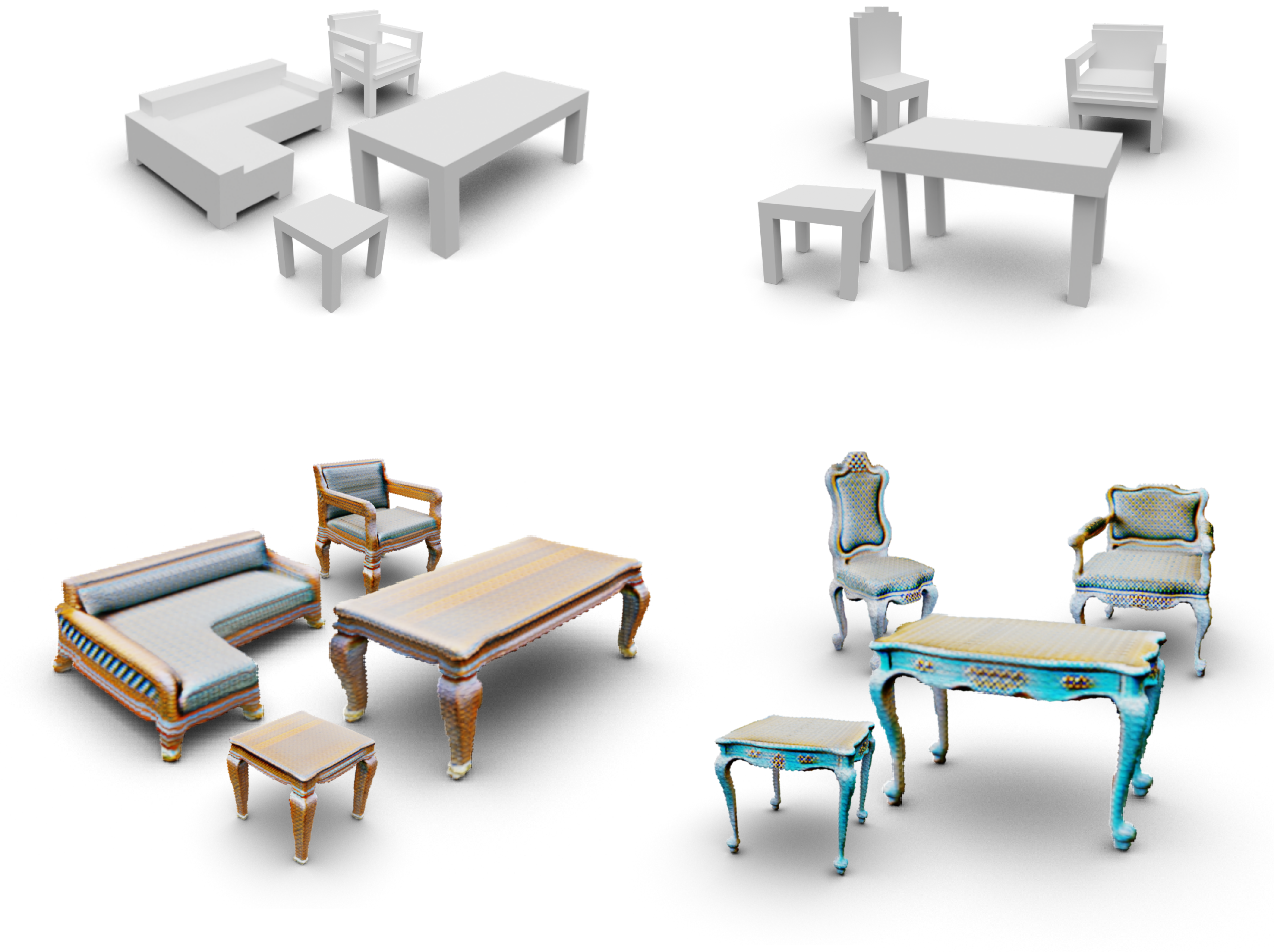}}

  \put(83, 230){\textit{``a classic furniture piece made of}}
  \put(80, 220){\textit{polished wood with subtle details''}}

  \put(360, 230){\textit{``an old Queen Anne style}}
  \put(350, 220){\textit{furniture in a light, neutral tone''}}

\end{picture}
\vspace{-10mm}
  \caption{Additional results of cross-category detailization. We show a collection of coarse voxels from the chair, table, couch, and bed classes on the top and the detailed shapes on the bottom. Our method can generate structurally varying shapes spanning multiple furniture categories with a consistent style.}
  \label{fig:additional_app_results}
\end{figure*}

Figure~\ref{fig:additional_app_results} shows additional results of cross-category detailization.

\clearpage

\begin{table}[ht]
    \begin{center}
    \caption{Quantitative comparison of structural guided geneartion on text prompt \textit{"a farmhouse chair with a cross-back design and a natural wood finish"}.
    }
    \label{tab:farmhouse_chair}
    \vspace{-3mm}
    \resizebox{1.0\columnwidth}{!}{
    \begin{tabular}{lcccc}
    \toprule
        & Render-FID $\downarrow$ & CLIP score $\uparrow$ & Strict-IoU $\uparrow$ & Loose-IoU $\uparrow$ \\
    \midrule
       ShaDDR & 194.624 & 21.523 & 0.511 & 0.636 \\
       Coin3D & 202.395 & 20.413 & 0.479 & 0.628 \\
       CLAY & 178.604 & 23.634 & \textbf{0.546} & 0.647 \\
       Ours & \textbf{176.069} & \textbf{24.758} & 0.510 & \textbf{0.651} \\
    \bottomrule
    \end{tabular}
    }
    \end{center}
\end{table}
\begin{table}[ht]
    \begin{center}
    \caption{Quantitative comparison of structural guided geneartion on text prompt \textit{"a Queen Anne chair with a leather back in a light, neutral tone and a cushioned seat"}.
    }
    \label{tab:queen_anne_chair}
    \vspace{-3mm}
    \resizebox{1.0\columnwidth}{!}{
    \begin{tabular}{lcccc}
    \toprule
        & Render-FID $\downarrow$ & CLIP score $\uparrow$ & Strict-IoU $\uparrow$ & Loose-IoU $\uparrow$ \\
    \midrule
       ShaDDR & 204.813 & 26.591 & 0.563 & 0.611 \\
       Coin3D & 218.345 & 19.371 & 0.617 & 0.694 \\
       CLAY & 185.332 & 24.182 & 0.673 & \textbf{0.761} \\
       Ours & \textbf{163.286} & \textbf{28.737} & \textbf{0.682} & 0.751 \\
    \bottomrule
    \end{tabular}
    }
    \end{center}
\end{table}
\begin{table}[ht]
    \begin{center}
    \caption{Quantitative comparison of structural guided geneartion on text prompt \textit{"a metal chair with a leather back and a cushioned seat"}.
    }
    \label{tab:metal_chair}
    \vspace{-3mm}
    \resizebox{1.0\columnwidth}{!}{
    \begin{tabular}{lcccc}
    \toprule
        & Render-FID $\downarrow$ & CLIP score $\uparrow$ & Strict-IoU $\uparrow$ & Loose-IoU $\uparrow$ \\
    \midrule
       ShaDDR & 183.891 & 22.385 & 0.507 & 0.594 \\
       Coin3D & 191.389 & 20.812 & 0.496 & 0.598 \\
       CLAY & 177.283 & 23.226 & 0.524 & 0.618 \\
       Ours & \textbf{173.923} & \textbf{23.239} & \textbf{0.578} & \textbf{0.682} \\
    \bottomrule
    \end{tabular}
    }
    \end{center}
\end{table}
\begin{table}[ht]
    \begin{center}
    \caption{Quantitative comparison of structural guided geneartion on text prompt \textit{"a traditional Japanese palace with tiled roofs and wooden walls"}.
    }
    \label{tab:japanese_palace}
    \vspace{-3mm}
    \resizebox{1.0\columnwidth}{!}{
    \begin{tabular}{lcccc}
    \toprule
        & Render-FID $\downarrow$ & CLIP score $\uparrow$ & Strict-IoU $\uparrow$ & Loose-IoU $\uparrow$ \\
    \midrule
       ShaDDR & 186.036 & 23.723 & 0.662 & 0.747 \\
       Coin3D & 218.126 & 21.283 & 0.643 & 0.737 \\
       CLAY & 172.284 & 25.283 & \textbf{0.684} & \textbf{0.792} \\
       Ours & \textbf{165.835} & \textbf{27.264} & 0.671 & 0.763 \\
    \bottomrule
    \end{tabular}
    }
    \end{center}
\end{table}
\begin{table}[ht]
    \begin{center}
    \caption{Quantitative comparison of structural guided geneartion on text prompt \textit{"a Gothic church with spires, detailed carvings, and flying buttresses"}.
    }
    \label{tab:gothic_church}
    \vspace{-3mm}
    \resizebox{1.0\columnwidth}{!}{
    \begin{tabular}{lcccc}
    \toprule
        & Render-FID $\downarrow$ & CLIP score $\uparrow$ & Strict-IoU $\uparrow$ & Loose-IoU $\uparrow$ \\
    \midrule
       ShaDDR & 179.472 & 23.623 & \textbf{0.745} & \textbf{0.817} \\
       Coin3D & 192.268 & 20.178 & 0.639 & 0.751 \\
       CLAY & \textbf{170.237} & 23.825 & 0.692 & 0.763 \\
       Ours & 171.235 & \textbf{25.836} & 0.729 & 0.797 \\
    \bottomrule
    \end{tabular}
    }
    \end{center}
\end{table}
\begin{table}[ht]
    \begin{center}
    \caption{Quantitative comparison of structural guided geneartion on text prompt \textit{"a Russian cathedral with domes and tall spires"}.
    }
    \label{tab:russian_cathedral}
    \vspace{-3mm}
    \resizebox{1.0\columnwidth}{!}{
    \begin{tabular}{lcccc}
    \toprule
        & Render-FID $\downarrow$ & CLIP score $\uparrow$ & Strict-IoU $\uparrow$ & Loose-IoU $\uparrow$ \\
    \midrule
       ShaDDR & 188.492 & 20.533 & \textbf{0.729} & \textbf{0.821} \\
       Coin3D & 194.126 & 19.175 & 0.621 & 0.756 \\
       CLAY & \textbf{170.003} & 21.528 & 0.711 & 0.819 \\
       Ours & 172.946 & \textbf{23.730} & 0.712 & 0.803 \\
    \bottomrule
    \end{tabular}
    }
    \end{center}
\end{table}
\begin{table}[ht]
    \begin{center}
    \caption{Quantitative comparison of structural guided geneartion on text prompt \textit{"a cake with chocolate dripping down the sides"}.
    }
    \label{tab:chocolate_cake}
    \vspace{-3mm}
    \resizebox{1.0\columnwidth}{!}{
    \begin{tabular}{lcccc}
    \toprule
        & Render-FID $\downarrow$ & CLIP score $\uparrow$ & Strict-IoU $\uparrow$ & Loose-IoU $\uparrow$ \\
    \midrule
       ShaDDR & 189.471 & 20.524 & 0.647 & 0.767 \\
       Coin3D & 197.271 & 18.274 & 0.529 & 0.636 \\
       CLAY & \textbf{172.194} & 24.482 & 0.612 & 0.728 \\
       Ours & 178.836 & \textbf{24.624} & \textbf{0.658} & \textbf{0.785} \\
    \bottomrule
    \end{tabular}
    }
    \end{center}
\end{table}
\begin{table}[ht]
    \begin{center}
    \caption{Quantitative comparison of structural guided geneartion on text prompt \textit{"a chair with a backrest shaped like a large maple leaf"}.
    }
    \label{tab:maple_leaf_chair}
    \vspace{-3mm}
    \resizebox{1.0\columnwidth}{!}{
    \begin{tabular}{lcccc}
    \toprule
        & Render-FID $\downarrow$ & CLIP score $\uparrow$ & Strict-IoU $\uparrow$ & Loose-IoU $\uparrow$ \\
    \midrule
       ShaDDR & 178.345 & 27.734 & 0.538 & 0.659 \\
       Coin3D & 193.561 & 22.491 & 0.510 & 0.627 \\
       CLAY & 177.284 & 25.381 & 0.564 & 0.668 \\
       Ours & \textbf{171.491} & \textbf{30.184} & \textbf{0.571} & \textbf{0.682} \\
    \bottomrule
    \end{tabular}
    }
    \end{center}
\end{table}

\clearpage
\begin{table}[ht]
    \begin{center}
    \caption{Ablation results on text prompt \textit{"a farmhouse chair with a cross-back design and a natural wood finish"}.
    }
    \label{tab:farmhouse_chair_ablation}
    \vspace{-3mm}
    \resizebox{1.0\columnwidth}{!}{
    \begin{tabular}{lcccc}
    \toprule
        & Render-FID $\downarrow$ & CLIP score $\uparrow$ & Strict-IoU $\uparrow$ & Loose-IoU $\uparrow$ \\
    \midrule
       $\lambda_{mask} = 0$ & 176.304 & 23.634 & 0.483 & 0.627 \\
       $\lambda_{mask} = 10^2$ & 177.936 & 24.237 & 0.489 & 0.631 \\
       $\lambda_{mask} = 10^3$ & 176.382 & 23.532 & 0.507 & 0.648 \\
       $\lambda_{mask} = 10^4$ & \textbf{176.069} & \textbf{24.758} & \textbf{0.510} & \textbf{0.651} \\
    \bottomrule
    \end{tabular}
    }
    \end{center}
\end{table}
\begin{table}[ht]
    \begin{center}
    \caption{Ablation results on text prompt \textit{"a Queen Anne chair with a leather back in a light, neutral tone and a cushioned seat"}.
    }
    \label{tab:queen_anne_chair_ablation}
    \vspace{-3mm}
    \resizebox{1.0\columnwidth}{!}{
    \begin{tabular}{lcccc}
    \toprule
        & Render-FID $\downarrow$ & CLIP score $\uparrow$ & Strict-IoU $\uparrow$ & Loose-IoU $\uparrow$ \\
    \midrule
       $\lambda_{mask} = 0$ & \textbf{163.022} & 28.281 & 0.652 & 0.729 \\
       $\lambda_{mask} = 10^2$ & 163.381 & 27.769 & 0.658 & 0.730 \\
       $\lambda_{mask} = 10^3$ & 163.836 & 28.172 & 0.676 & 0.747 \\
       $\lambda_{mask} = 10^4$ & 163.286 & \textbf{28.737} & \textbf{0.682} & \textbf{0.751} \\
    \bottomrule
    \end{tabular}
    }
    \end{center}
\end{table}
\begin{table}[ht]
    \begin{center}
    \caption{Ablation results on text prompt \textit{"a metal chair with a leather back and a cushioned seat"}.
    }
    \label{tab:metal_chair_ablation}
    \vspace{-3mm}
    \resizebox{1.0\columnwidth}{!}{
    \begin{tabular}{lcccc}
    \toprule
        & Render-FID $\downarrow$ & CLIP score $\uparrow$ & Strict-IoU $\uparrow$ & Loose-IoU $\uparrow$ \\
    \midrule
       $\lambda_{mask} = 0$ & 174.197 & \textbf{23.625} & 0.518 & 0.633 \\
       $\lambda_{mask} = 10^2$ & 173.823 & 23.116 & 0.522 & 0.637 \\
       $\lambda_{mask} = 10^3$ & \textbf{173.468} & 23.361 & 0.569 & 0.663 \\
       $\lambda_{mask} = 10^4$ & 173.923 & 23.239 & \textbf{0.578} & \textbf{0.682} \\
    \bottomrule
    \end{tabular}
    }
    \end{center}
\end{table}
\begin{table}[ht]
    \begin{center}
    \caption{Ablation results on text prompt \textit{"a traditional Japanese palace with tiled roofs and wooden walls"}.
    }
    \label{tab:japanese_palace_ablation}
    \vspace{-3mm}
    \resizebox{1.0\columnwidth}{!}{
    \begin{tabular}{lcccc}
    \toprule
        & Render-FID $\downarrow$ & CLIP score $\uparrow$ & Strict-IoU $\uparrow$ & Loose-IoU $\uparrow$ \\
    \midrule
       $\lambda_{mask} = 0$ & 165.963 & 26.927 & 0.611 & 0.706 \\
       $\lambda_{mask} = 10^2$ & 165.782 & 26.694 & 0.609 & 0.694 \\
       $\lambda_{mask} = 10^3$ & \textbf{165.196} & 26.379 & 0.652 & 0.757 \\
       $\lambda_{mask} = 10^4$ & 165.835 & \textbf{27.264} & \textbf{0.671} & \textbf{0.763} \\
    \bottomrule
    \end{tabular}
    }
    \end{center}
\end{table}
\begin{table}[ht]
    \begin{center}
    \caption{Ablation results on text prompt \textit{"a Gothic church with spires, detailed carvings, and flying buttresses"}.
    }
    \label{tab:gothic_church_ablation}
    \vspace{-3mm}
    \resizebox{1.0\columnwidth}{!}{
    \begin{tabular}{lcccc}
    \toprule
        & Render-FID $\downarrow$ & CLIP score $\uparrow$ & Strict-IoU $\uparrow$ & Loose-IoU $\uparrow$ \\
    \midrule
       $\lambda_{mask} = 0$ & 171.397 & \textbf{25.913} & 0.658 & 0.746 \\
       $\lambda_{mask} = 10^2$ & \textbf{170.731} & 25.612 & 0.633 & 0.739 \\
       $\lambda_{mask} = 10^3$ & 171.169 & 25.081 & 0.689 & 0.784 \\
       $\lambda_{mask} = 10^4$ & 171.235 & 25.836 & \textbf{0.729} & \textbf{0.797} \\
    \bottomrule
    \end{tabular}
    }
    \end{center}
\end{table}
\begin{table}[ht]
    \begin{center}
    \caption{Ablation results on text prompt \textit{"a Russian cathedral with domes and tall spires"}.
    }
    \label{tab:russian_cathedral_ablation}
    \vspace{-3mm}
    \resizebox{1.0\columnwidth}{!}{
    \begin{tabular}{lcccc}
    \toprule
        & Render-FID $\downarrow$ & CLIP score $\uparrow$ & Strict-IoU $\uparrow$ & Loose-IoU $\uparrow$ \\
    \midrule
       $\lambda_{mask} = 0$ & \textbf{171.291} & 23.172 & 0.612 & 0.728 \\
       $\lambda_{mask} = 10^2$ & 172.735 & 23.561 & 0.633 & 0.747 \\
       $\lambda_{mask} = 10^3$ & 173.013 & 23.597 & 0.684 & 0.793 \\
       $\lambda_{mask} = 10^4$ & 172.946 & \textbf{23.730} & \textbf{0.712} & \textbf{0.803} \\
    \bottomrule
    \end{tabular}
    }
    \end{center}
\end{table}
\begin{table}[ht]
    \begin{center}
    \caption{Ablation results on text prompt \textit{"a cake with chocolate dripping down the sides"}.
    }
    \label{tab:chocolate_cake_ablation}
    \vspace{-3mm}
    \resizebox{1.0\columnwidth}{!}{
    \begin{tabular}{lcccc}
    \toprule
        & Render-FID $\downarrow$ & CLIP score $\uparrow$ & Strict-IoU $\uparrow$ & Loose-IoU $\uparrow$ \\
    \midrule
       $\lambda_{mask} = 0$ & \textbf{177.304} & 23.812 & 0.609 & 0.716 \\
       $\lambda_{mask} = 10^2$ & 177.782 & 24.018 & 0.603 & 0.701 \\
       $\lambda_{mask} = 10^3$ & 178.423 & 24.184 & 0.643 & 0.769 \\
       $\lambda_{mask} = 10^4$ & 178.836 & \textbf{24.624} & \textbf{0.658} & \textbf{0.785} \\
    \bottomrule
    \end{tabular}
    }
    \end{center}
\end{table}
\begin{table}[ht]
    \begin{center}
    \caption{Ablation results on text prompt \textit{"a chair with a backrest shaped like a large maple leaf"}.
    }
    \label{tab:maple_leaf_chair_ablation}
    \vspace{-3mm}
    \resizebox{1.0\columnwidth}{!}{
    \begin{tabular}{lcccc}
    \toprule
        & Render-FID $\downarrow$ & CLIP score $\uparrow$ & Strict-IoU $\uparrow$ & Loose-IoU $\uparrow$ \\
    \midrule
       $\lambda_{mask} = 0$ & 171.622 & \textbf{30.383} & 0.533 & 0.659 \\
       $\lambda_{mask} = 10^2$ & 171.293 & 29.962 & 0.526 & 0.642 \\
       $\lambda_{mask} = 10^3$ & \textbf{170.932} & 29.075 & 0.566 & 0.663 \\
       $\lambda_{mask} = 10^4$ & 171.491 & 30.184 & \textbf{0.571} & \textbf{0.682} \\
    \bottomrule
    \end{tabular}
    }
    \end{center}
\end{table}

\clearpage

\clearpage

\clearpage

\clearpage

\clearpage










\end{document}